%% file: ICML_paper.tex
\documentclass{article}

\usepackage{times, cite, natbib} 
\usepackage{amsmath, amssymb, amsthm, bm} 
\usepackage{algorithm, algorithmic}
\usepackage{graphicx, subfigure} 
\usepackage{hyperref, enumitem}
\input{macro+variables}

\usepackage[accepted]{icml2016}

\icmltitlerunning{Compressive Spectral Clustering}

\begin{document} 

\twocolumn[
\icmltitle{Compressive Spectral Clustering}
\icmlauthor{Nicolas $\mbox{TREMBLAY}^{*\dag}$}{nicolas.tremblay@inria.fr}
\icmlauthor{Gilles $\mbox{PUY}^{\S *}$}{gilles.puy@technicolor.com}
\icmlauthor{R\'emi $\mbox{GRIBONVAL}^*$}{remi.gribonval@inria.fr}
\icmlauthor{Pierre $\mbox{VANDERGHEYNST}^{\dag *}$}{pierre.vandergheynst@epfl.ch}
\icmladdress{$\,^*$ INRIA Rennes - Bretagne Atlantique, Campus de Beaulieu, FR-35042 Rennes Cedex, France\\
            $~^\dag$ Institute of Electrical Engineering, Ecole Polytechnique F{\'e}d{\'e}rale de Lausanne (EPFL), CH-1015 Lausanne, Switzerland\\
            $~^\S$ Technicolor, 975 Avenue des Champs Blancs, 35576 Cesson-S\'{e}vign\'{e}, France}

\icmlkeywords{graph signal processing, spectral clustering}
\vskip 0.3in
]

\begin{abstract} 
Spectral clustering has become a popular technique due to its high performance in 
many contexts. It comprises three main steps: create a similarity graph between $N$ objects to cluster, 
compute the first $k$ eigenvectors of its Laplacian matrix to define a feature vector for each 
object, and run $k$-means on these features to separate objects into $\nbClass$ classes. Each of these three 
steps becomes computationally intensive for large $N$ and/or $k$. We propose to speed up the 
last two steps based on recent results in the emerging field of graph signal processing:   
graph filtering of random signals, and random sampling of bandlimited graph signals. We 
prove that our method, with a gain in computation time that can reach several orders of magnitude, 
is in fact an approximation of spectral clustering, for which we are able to control the 
error. We test the performance of our method  on artificial and 
real-world network data. 
\end{abstract} 
%


%
\section{Introduction}

Spectral clustering (SC) is a fundamental tool in 
data mining~\cite{Nascimento_Eur_J_Oper_Reas2011}. Given a set of $\nbVert$ data 
points $\{\vec{x}_1, \ldots, \vec{x}_N \}$, the goal is to partition this set into $k$ weakly inter-connected 
clusters. Several spectral clustering algorithms exist, 
\eg,~\cite{ng_ANIPS2002,shi_TPAMI2000,belkin2003laplacian,zelnik_NIPS2004}, but all follow the same 
scheme. 
First, compute weights $\ma{W}_{ij} \geq 0$ that model the similarity between pairs of data points 
$(\vec{x}_i, \vec{x}_j)$. This gives rise to a graph $\Graph$ with $N$ nodes and adjacency matrix 
$\ma{W} = (\ma{W}_{ij})_{1 \leq i,j \leq \nbVert} \in \Rbb^{\nbVert \times \nbVert}$. Second, compute 
the first $\nbClass$ eigenvectors $\ma{U}_k := (\fou_{1}, \ldots, \fou_{k}) \in \Rbb^{\nbVert \times k}$ 
of the Laplacian matrix  $\Lap \in \Rbb^{\nbVert \times \nbVert}$ associated to $\Graph$ (see Sec.~\ref{sec:gsp} for $\Lap$'s definition). And finally, run $k$-means using the rows of $\ma{U}_k$ as feature 
vectors to partition the $\nbVert$ data points into $k$ clusters. This $k$-way scheme is a generalisation 
of Fiedler's pioneering work~\yrcite{fiedler1973algebraic}. 

SC is mainly used in two contexts: $1)$ if the $\nbVert$ 
data points show particular structures (\eg, concentric circles) for which naive $k$-means clustering fails; 
$2$) if the input data is directly a graph $\Graph$ modeling a network~\cite{white2005spectral}, such as social, neuronal, or transportation networks.
SC suffers nevertheless from three main computational bottlenecks for large $\nbVert$ and/or $\nbClass$: 
the creation of the similarity matrix $\ma{W}$; the partial eigendecomposition of the 
graph Laplacian matrix $\Lap$; and $k$-means.

\subsection{Related work}

Circumventing these bottlenecks has raised a significant interest in the past decade. 
Several authors have proposed ideas to tackle the eigendecomposition bottleneck, \eg, via the power 
method~\cite{Boutsidis_ICML2015,lin_ICML2010}, via a careful optimisation of diagonalisation algorithms 
in the context of SC~\cite{Liu_AIR2007}, or via matrix column-subsampling such as in the Nystr\"om 
method~\cite{Fowlkes_TPAMI2004}, the nSPEC and cSPEC methods of~\cite{Wang_AKDDM2009}, or 
in~\cite{chen2011large,Sakai_MLDMPR2009}. All these methods aim to quickly compute feature vectors, 
but $k$-means is still applied on $\nbVert$ feature vectors. Other authors, inspired by research aiming at reducing k-means complexity~\cite{Jain_PRL2010}, such 
as the line of work on coresets~\cite{har2004coresets}, have proposed to 
circumvent $k$-means 
in high dimension by subsampling a few data points out of the $N$ available ones, 
applying SC on its reduced similarity graph, and 
interpolating the results back on the 
complete dataset. One can find similar methods in~\cite{Yan_KDD2009} and~\cite{Wang_AKDDM2009}'s eSPEC proposition, where two 
different interpolation methods are used. Both methods are heuristic: there is no proof that these methods 
approach the results of SC. 
Also, let us mention~\cite{dhillon_TPAMI2007} that circumvents 
both the eigendecomposition and the $k$-means bottlenecks: the authors reduce 
the graph's size by successive aggregation of nodes, apply SC on this small graph, and 
propagate the results on the complete graph using kernel $k$-means to control interpolation errors. 
The kernel is computed so that kernel $k$-means and SC share the same objective 
function~\cite{Filippone_PR2008}. Finally, we mention 
works~\cite{boutsidis2011stochastic,cohen2015dimensionality} that concentrate on reducing the feature vectors' 
dimension in the $k$-means problem, but do not sidestep the eigendecomposition nor 
the large $N$ issues.

\subsection{Contribution: compressive clustering}

In this work, inspired by recent advances in the emerging field of graph signal 
processing~\cite{shuman_SPMAG2013, sandryhaila_SPMAG2014}, 
we circumvent SC's last two bottlenecks and detail a fast approximate spectral clustering method 
for large datasets, as well as the supporting theory. 
We suppose that the Laplacian matrix $\ma{L} \in \Rbb^{\nbVert \times \nbVert}$ of $\Graph$ is given. Our method 
is made of two ingredients. 

The first ingredient builds upon recent works~\cite{tremblay_puy_icassp16, NIPS_sidestepping_svd} that 
avoid the costly computation of the eigenvectors 
of $\Lap$ by filtering $O(\log(k))$ random signals on $\Graph$ that will then serve as feature vectors 
to perform clustering. We show in this paper how to incorporate the effects of non-ideal, but computationally 
efficient, graph filters on the quality of the feature vectors used for clustering.

The second ingredient uses a recent sampling theory of bandlimited graph-signals~\cite{Puy_ARXIV2015} to reduce 
the computational complexity of $k$-means. Using the fact that the indicator vectors of each cluster 
are approximately bandlimited on $\Graph$, we prove that clustering a random subset of $O(k\log(k))$ nodes 
of $\Graph$ using random features vectors of size $O(\log(k))$ is sufficient to infer rapidly and accurately 
the cluster label of all $\nbVert$ nodes of the graph. Note that the complexity of $k$-means is reduced to 
$O(k^2\log^2(k))$ instead of $O(Nk^2)$ for SC. One readily sees that this method scales easily 
to large datasets, as will be demonstrated on artifical and real-world datasets containing up to 
$\nbVert=10^6$ nodes. 

The proposed compressive spectral clustering method can be summarised as follows:
\begin{itemize}[noitemsep, nolistsep]
\item generate a feature vector for each node by filtering $O(\log(k))$ random Gaussian signals on $\Graph$;
\item sample $O(k\log(k))$ nodes from the full set of nodes; 
\item cluster the reduced set of nodes; 
\item interpolate the cluster indicator vectors back to the complete graph.
\end{itemize}
%


%
\section{Background}
\label{sec:gsp}

\subsection{Graph signal processing}
Let $\mathcal{G}=(\mathcal{V},\mathcal{E},\mathbf{W})$ be an undirected weighted 
graph with $\mathcal{V}$ the set 
of $\nbVert$ nodes, $\mathcal{E}$ the set of edges, and $\mathbf{W}$ the weighted adjacency 
matrix such that $W_{ij}=W_{ji}\geq0$ is 
the weight of the edge between nodes $i$ and $j$.  

\textbf{The graph Fourier matrix.}
Consider the graph's normalized  Laplacian 
matrix $\Lap=\ma{I}-\ma{D}^{-1/2}\ma{W}\ma{D}^{-1/2}$ where $\ma{I}$ is the identity in dimension $\nbVert$, and 
$\ma{D}$ is diagonal  
with $\ma{D}_{ii}= \sum_{j\neq i} \ma{W}_{ij}$. 
$\Lap$ is real symmetric and positive semi-definite, therefore diagonalizable as $\Lap=\Fou\Lambda\Fou^\adjoint$, where 
$\Fou :=\left(\fou_1|\fou_2|\dots|\fou_\nbVert\right) \in \Rbb^{\nbVert \times \nbVert}$ is the orthonormal basis of eigenvectors and 
$\ma{\Lambda} \in \Rbb^{\nbVert \times \nbVert}$ the diagonal matrix containing its sorted eigenvalues : 
$0=\lambda_1\leq\dots\leq\lambda_{\nbVert}\leq2$~\cite{chung_book1997}. 
By analogy to the continuous Laplacian operator whose eigenfunctions
are the classical Fourier modes and eigenvalues their 
squared frequencies, the columns of $\Fou$ are considered as the graph's 
Fourier modes, and 
$\{\sqrt{\lambda_l}\}_{l}$ as its set of associated ``frequencies''~\cite{shuman_SPMAG2013}. 
Other types of graph Fourier matrices have been proposed, \eg, \cite{sandryhaila_TSP2013}, 
but in order to exhibit the link between graph signal processing and SC, the Laplacian-based Fourier matrix appears more natural.

\textbf{Graph filtering.}
The graph Fourier transform $\hat{\vec{x}}$ of a signal $\vec{x}$ defined 
on the nodes of the graph (called a \textit{graph signal}) reads: $\hat{\vec{x}}=\Fou^\adjoint \vec{x}$. 
Given a continuous filter function $h$ defined on $[0,2]$, its 
associated graph filter 
operator $\ma{H}\in\Rbb^{\nbVert\times \nbVert}$ is defined as
 $\ma{H} := h(\Lap) = \Fou h(\Eig) \Fou^\adjoint$,
where $h(\Eig) := \mbox{diag}(h(\lambda_1),h(\lambda_2),\cdots,h(\lambda_\nbVert))$. The signal $\vec{x}$ filtered by $h$ is $\ma{H}\vec{x}$. 
In the following, we consider ideal low-pass filters, denoted by $h_{\lambda_c}$, 
that satisfy, for all $\lambda \in [0, 2]$,
\begin{align}
 \label{eq:ideal}
  h_{\lambda_c}(\lambda)=1, \mbox{ if } \lambda \leq \lambda_c, 
  \; \mbox{ and } \;
  h_{\lambda_c}(\lambda)=0, \mbox{ if not.}
\end{align}
Denote by $\ma{H}_{\lambda_c}$ the graph filter operator associated to $h_{\lambda_c}$. 

\textbf{Fast graph filtering.}
\label{subsec:fast_filt}
To filter a signal by $h$ without diagonalizing $\Lap$, one may approximate $h$ by a polynomial 
$\tilde{h}$ of order $\ordPA$ satisfying
%
$\tilde{h}(\lambda) := \sum_{l=0}^\ordPA\alpha_l\lambda^l\simeq h(\lambda)$ 
%
for all $\lambda\in[0,2]$, 
where $\alpha_1, \ldots, \alpha_\ordPA \in \Rbb$. In matrix form, we have
%
$
\ma{\tilde{H}} := \tilde{h}(\Lap) = \sum_{l=0}^\ordPA\alpha_l\Lap^l\simeq\ma{H}.
$
%
\emph{Let us highlight that we never compute the potentially dense matrix $\ma{\tilde{H}}$ in practice}. Indeed, 
we are only    
interested in the result of the filtering operation: 
$\ma{\tilde{H}}\vec{x} = \sum_{l=0}^\ordPA\alpha_l\Lap^l\vec{x} \approx \ma{H}\vec{x}$ for 
$\vec{x} \in \Rbb^\nbVert$, obtainable with only $\ordPA$ successive 
matrix-vector multiplications with $L$. The computational complexity of filtering a signal is thus $O(\ordPA\#\mathcal{E})$, 
where $\#\mathcal{E}$ is the number of edges of $\Graph$. 

\begin{algorithm}[tb]
   \caption{Spectral Clustering~\cite{ng_ANIPS2002}}
   \label{alg:SC}
\begin{algorithmic}
\STATE \textbf{Input:} The Laplacian matrix $\Lap$, the number of clusters $k$
%
  \STATE $\bm{1\cdot}$ Compute $\Fou_\nbClass\in\Rbb^{\nbVert\times\nbClass}$, $\Lap$'s first $\nbClass$ 
 eigenvectors: $\Fou_\nbClass=\left(\fou_1|\fou_2|\cdots|\fou_\nbClass\right)$. 
 \STATE $\bm{2\cdot}$ Form the matrix $\ma{Y}_\nbClass\in\mathbb{R}^{\nbVert\times \nbClass}$ from $\Fou_\nbClass$ by 
 normalizing each of $\Fou_\nbClass$'s
 rows to unit length: $\left(\ma{Y}_{\nbClass}\right)_{ij}=\left(\Fou_{\nbClass}\right)_{ij}/\sqrt{\sum_{j=1}^\nbClass \Fou_{ij}^2}$.
  \STATE $\bm{3\cdot}$ Treat each node $i$ as a point in $\Rbb^\nbClass$ by defining its feature 
 vector $\feat_i\in\Rbb^\nbClass$ 
 as the transposed $i$-th row of $\ma{Y}_\nbClass$:
 \begin{equation*}
  \feat_i := \ma{Y}_\nbClass^\adjoint\vec{\delta}_i,
 \end{equation*}
where $\delta_i(j)=1$ if $j=i$ and $0$ otherwise.
  \STATE $\bm{4\cdot}$ To obtain 
 $\nbClass$ clusters, run $k$-means with the Euclidean distance: 
 \begin{equation}
\label{eq:Dist2est}
 D_{ij} :=\norm{\feat_i-\feat_j}
\end{equation}
\end{algorithmic}
\end{algorithm}

\subsection{Spectral clustering}
\label{subsec:SP}
We choose here Ng et al.'s method~\yrcite{ng_ANIPS2002} based on the normalized 
Laplacian as our standard SC method. 
The input is the adjacency matrix $\ma{W}$ representing the pairwise similarity 
of all the $\nbVert$ objects to cluster\footnote{In network analysis, the raw data is directly 
$\ma{W}$. In the case where one starts with a
set of data points $(\vec{x}_1,\ldots,\vec{x}_\nbVert)$, the first step 
consists in deriving $\ma{W}$ from the pairwise similarities $s(\vec{x}_i,\vec{x}_j)$. 
See~\cite{vonluxburg_StatComp2011} for several choices of similarity 
 measure $s$ 
 and several ways to create $\ma{W}$ from the $s(\vec{x}_i,\vec{x}_j)$.}. 
After computing its Laplacian $\Lap$, follow Alg.~\ref{alg:SC} to find $\nbClass$ classes.


%
\section{Principles of CSC}

Compressive spectral clustering (CSC) circumvents two of SC's bottlenecks, the partial diagonalisation of the Laplacian 
 and the 
high-dimensional $k$-means, thanks to the following ideas.

1) Perform a  controlled estimation $\tilde{D}_{ij}$ of the spectral clustering distance $D_{ij}$ (see Eq~\eqref{eq:Dist2est}), 
\textit{without} partially diagonalizing the Laplacian, 
by fast filtering a few random signals with the polynomial approximation $\tilde{h}_{\eig_{\nbClass}}$ of
the ideal low pass filter $h_{\eig_{\nbClass}}$ (see Eq.~\eqref{eq:ideal}). 
A theorem recently published independently by two 
teams~\cite{tremblay_puy_icassp16, NIPS_sidestepping_svd} shows that this is possible when 
there is no normalisation step (step 2 in Alg.~\ref{alg:SC}) and 
when 
the order $p$ of the polynomial approximation tends to infinity, \ie, when 
$\tilde{h}_{\eig_{\nbClass}}=h_{\eig_{\nbClass}}$. 
In Sec.~\ref{subsec:ideal_filtering}, we provide a first extension of this theorem that takes into account normalisation. A complete extension that also takes into account the polynomial approximation 
error is presented in Sec.~\ref{subsec:non_ideal_filtering}.
 
2) Run $k$-means on $\nbVertRed$ randomly selected feature vectors out of the $\nbVert$ available ones 
- thus clustering the corresponding $\nbVertRed$ nodes into $\nbClass$ groups - 
and interpolate the result back on the full graph. To guarantee robust reconstruction, we take advantage 
of our recent results on random sampling of $k$-bandlimited graph signals. 
In Sec.~\ref{subsec:ideal_sampling}, we explain why these 
results are applicable to clustering and show that it is sufficient to sample $n=O(\nbClass\log{\nbClass})$ 
features only! Note that to cluster data into $k$ groups, one needs \textit{at least} $k$ samples. 
This result is thus optimal up to the extra $\log{\nbClass}$ factor.

\subsection{Ideal filtering of random signals}
\label{subsec:ideal_filtering}

\begin{definition}[Local cumulative coherence]
Given a graph $\Graph$, the local cumulative coherence of order $\nbClass$ at node $i$ is\footnote{Throughout this paper, $\norm{.}$ stands for the usual $\ell_2$-norm.}
$v_{\nbClass} (i) :=  \norm{\Fou_\nbClass^\adjoint \vec{\delta}_i}=\sqrt{\sum_{j=1}^\nbClass \Fou_{ij}^2}$.
\end{definition}
Let us define the diagonal matrix: $\ma{V}_\nbClass(i,i)=1/v_{\nbClass} (i)$. Note that we assume that $v_{\nbClass} (i) > 0$. Indeed, in the pathologic cases where $v_{\nbClass} (i) = 0$ for some nodes $i$, 
step 2 of the standard SC algorithm cannot be run either. Now, consider the matrix $\ma{R}=\left(\vec{r}_1|\rv_2|\cdots|\rv_\nbFeat\right)\in\Rbb^{\nbVert\times\nbFeat}$ 
consisting of $\nbFeat$ 
random signals $\rv_i$, whose components are independent Bernouilli, Gaussian, or sparse (as in Theorem 1.1 of~\cite{Achlioptas_JCSS2003}) 
random variables. To fix ideas in the following, we consider the components as independent random Gaussian 
variables of mean zero and  
variance $1/\nbFeat$. 
Consider the coherence-normalized filtered version of $\RV$, $\ma{V}_\nbClass\ma{H}_{\eig_{\nbClass}}\RV\in\Rbb^{\nbVert\times\nbFeat}$, and 
define node $i$'s new feature vector $\tilde{\feat}_i\in\Rbb^\nbFeat$ as the transposed $i$-th line of this filtered matrix: 
\begin{equation*}
 \tilde{\feat}_i := (\ma{V}_\nbClass\ma{H}_{\eig_{\nbClass}}\RV)^\adjoint\vec{\delta}_i.
\end{equation*}
The following theorem shows that, for large enough $\nbFeat$, 
\begin{equation*}
 \tilde{D}_{ij} := \norm{\tilde{\feat}_i-\tilde{\feat}_j}=\norm{(\ma{V}_\nbClass\ma{H}_{\eig_{\nbClass}}\RV)^\adjoint(\vec{\delta}_i-\vec{\delta}_j)}
\end{equation*}
is a good estimation of $D_{ij}$ with high probability. 

\begin{theorem}
\label{th:id_case_filt}
Let $\epsilon\in]0,1]$ and $\beta>0$ be given. If $\nbFeat$ is larger than
\begin{equation*}
 \frac{4+2\beta}{\epsilon^2/2-\epsilon^3/3}\log{\nbVert},
\end{equation*}
then with probability at least
  $1-\nbVert^{-\beta}$,
 we have
\begin{equation*}
(1-\epsilon)D_{ij}\leq \tilde{D}_{ij}\leq (1+\epsilon)D_{ij}.
\end{equation*}
for all $(i, j) \in \{1, \ldots, \nbVert \}^2$.
\end{theorem}
The proof is provided in the supplementary material. 

In Sec.~\ref{subsec:non_ideal_filtering}, we generalize this result to the real-world case 
where the low-pass filter is approximated by a finite order polynomial; we also prove that, as announced 
in the introduction, one only needs $\nbFeat=O(\log{\nbClass})$ features when using the downsampling scheme 
that we now detail.

\subsection{Downsampling and interpolation}
\label{subsec:ideal_sampling}

For $j=1, \ldots, \nbClass$, let us denote by $\indClus_j \in \Rbb^{\nbVert}$ the ground-truth indicator vector 
of cluster $\Clus_j$, \ie,
\begin{align*}
(\indClus_j)_i := 
\left\{
\begin{array}{ll}
1 & \text{if } i \in \Clus_j, \\
0 & \text{otherwise},
\end{array}
\right.
\quad \forall i \in \{1, \ldots, \nbVert \}.
\end{align*}
To estimate $\indClus_j$, one could run $k$-means on the $\nbVert$ feature vectors 
$\{\tilde{\feat}_{1}, \ldots, \tilde{\feat}_{\nbVert} \}$
, as done in~\cite{tremblay_puy_icassp16, NIPS_sidestepping_svd}. 
Yet, this is still inefficient for large $N$. 
To reduce the computational cost further, we propose to run $k$-means on a small subset of $\nbVertRed$ feature vectors only. 
The goal is then to infer the labels of all $\nbVert$ nodes from the labels of the $\nbVertRed$ sampled nodes. 
To this end, we need 1) a low-dimensional model that captures the regularity of the vectors $\indClus_j$, 
2) to make sure that enough information is preserved after sampling to be able to recover the 
vectors $\indClus_j$, and 3) an algorithm that rapidly and accurately estimates the vectors $\indClus_j$ 
by exploiting their regularity.

\subsubsection{The low-dimensional model}

For a simple regular (with nodes of same degree) graph of $k$ disconnected clusters, 
it is easy to check that $\{ \indClus_1, \ldots, \indClus_k \}$ form a set of orthogonal 
eigenvectors of $\Lap$ with eigenvalue $0$. All indicator vectors $\indClus_j$ therefore live 
in $\spann{(\Fou_\nbClass)}$. For general graphs, we assume that the indicator vectors $\indClus_j$ 
live close to $\spann{(\Fou_\nbClass)}$, \ie, the difference between any $\indClus_j$ and its orthogonal 
projection onto $\spann(\Fou_\nbClass)$ is small. Experiments in Section \ref{sec:results} will 
confirm that it is a good enough model to recover the cluster indicator vectors.

In graph signal processing words, one can say that $\indClus_j$ is approximately $\nbClass$-bandlimited, 
\ie, its $\nbClass$ first graph Fourier coefficients bear most of its energy. There has been recently a surge 
of interest around adapting classical sampling theorems to such bandlimited graph 
signals~\cite{chen_TSP2015,anis_arxiv2015_long,Tsitsvero_ARXIV2015,Marques_TSP2016}. We rely here on the 
random sampling strategy proposed in~\cite{Puy_ARXIV2015} to select a subset of $n$ nodes.

\subsubsection{Sampling and interpolation}
The subset of feature vectors is selected by drawing $\nbVertRed$ indices 
$\Omega := \{\omega_1, \ldots, \omega_\nbVertRed\}$ uniformly at random from  
$\{1, \ldots, \nbVert \}$ without replacement. Running $k$-means on the subset of features 
$\{\tilde{\feat}_{\omega_1}, \ldots,  \tilde{\feat}_{\omega_\nbVertRed} \}$ 
thus yields a clustering of the $\nbVertRed$ sampled nodes into $\nbClass$ clusters. 
We denote by $\indClus_j^r \in \Rbb^{\nbVertRed}$ the resulting low-dimensional indicator vectors. 
Our goal is now to recover $\indClus_j$ from $\indClus_j^r$.

Consider that $k$-means is able to correctly identify $\indClus_1, \ldots, \indClus_\nbClass \in \Rbb^\nbVert$ 
 using the original set of features $\{\feat_{1}, \ldots,  \feat_{\nbVert} \}$ with the SC algorithm 
 (otherwise, CSC is doomed to fail from the start). Results 
 in~\cite{tremblay_puy_icassp16, NIPS_sidestepping_svd}
show that $k$-means is also able to identify the clusters using the feature vectors 
$\{\tilde{\feat}_{1}, \ldots,  \tilde{\feat}_{\nbVert} \}$. This is explained theoretically by the fact that 
the distance between all pairs of feature vectors is preserved (see Theorem~\ref{th:id_case_filt}). Then, as 
choosing a subset $\{\tilde{\feat}_{\omega_1}, \ldots,  \tilde{\feat}_{\omega_\nbVertRed} \}$ 
of $\{\tilde{\feat}_{1}, \ldots,  \tilde{\feat}_{\nbVert} \}$ does not change the distance between the 
feature vectors, we can hope that $k$-means correctly clusters the $\nbVertRed$ sampled nodes, provided 
that each cluster is sufficiently sampled. Experiments in Sec.~\ref{sec:results} will confirm this 
intuition. In this ideal situation, we have 
\begin{equation}
\label{eq:hypo}
\indClus_j^r  = \Meas \, \indClus_j,
\end{equation}
where 
$\Meas \in \Rbb^{\nbVertRed \times \nbVert}$ is the sampling matrix satisfying:
\begin{align}
\label{eq:subsampling_matrix_def}
\Meas_{ij} := 
\left\{
\begin{array}{ll}
1 & \text{if } j = \omega_i, \\
0 & \text{otherwise}.
\end{array}
\right.
\end{align}
To recover $\indClus_j$ from its $\nbVertRed$ observations $\indClus_j^r$, Puy et al.~\yrcite{Puy_ARXIV2015} 
show that the solution to the optimisation problem 
\begin{align}
\label{eq:dec}
\min_{\sig \in \Rbb^\nbVert} \norm{\Meas \sig - \indClus_j^r}_2^2 + \reg \; \sig^\adjoint g(\Lap) \sig,
\end{align}
is a faithful~\footnote{precise error bounds are provided in~\cite{Puy_ARXIV2015}.} estimation of $\indClus_j$, provided that $\indClus_j$ is close to $\spann(\Fou_k)$ and 
that $\Meas$ satisfies the restricted isometry property (discussed in the next subsection). 
In \refeq{eq:dec}, $\reg>0$ is a regularisation parameter and $g$ a positive non-decreasing polynomial 
function (see Section \ref{subsec:fast_filt} for the definition of $g(\Lap)$). This reconstruction scheme  
is proved to be robust to: 1)~observation noise, \ie, to imperfect clustering of the $\nbVertRed$ nodes in our 
context; 2)~ model errors, \ie, the indicator vectors do not need to be exactly in $\spann{(\Fou_\nbClass)}$ 
for the method to work. Also, the performance is shown 
to depend on the ratio $g(\lambda_{\nbClass})/g(\eig_{\nbClass+1})$. The smaller it is, the better the 
reconstruction. To decrease this ratio, we decide to approximate the ideal high-pass filter 
$g_{\eig_{\nbClass}}(\lambda)=1-h_{\eig_{\nbClass}}(\lambda)$ for the reconstruction. Remark that this filter 
favors the recovery of signals living in $\spann{(\Fou_\nbClass)}$. The approximation 
$\tilde{g}_{\eig_{\nbClass}}$ of $g_{\eig_{\nbClass}}$ is obtained using a polynomial (as in 
Sec.~\ref{subsec:fast_filt}), which permits us to find fast algorithms to solve \refeq{eq:dec}.

\subsubsection{How many features to sample?}

We terminate this section by providing the theoretical number of features $\nbVertRed$ one needs to sample 
in order to make sure that the indicator vectors can be faithfully recovered. This number is driven by the following quantity.
\begin{definition}[Global cumulative coherence]
The global cumulative coherence of order $\nbClass$ of the graph $\Graph$ is
$
\cumCoh_{\nbClass}~:=~\sqrt{\nbVert}~\cdot~\max_{1 \leq i \leq \nbVert} \left\{ v_\nbClass(i) \right\}.
$
\end{definition}
It is shown in~\cite{Puy_ARXIV2015} that $\cumCoh_{\nbClass} \in [\nbClass^{1/2}, \nbVert^{1/2}]$. 
\begin{theorem}[\cite{Puy_ARXIV2015}]
Let $\Meas$ be a random sampling matrix constructed as in \refeq{eq:subsampling_matrix_def}. 
For any $\delta, \epsilon \in \;]0, 1[$,
\begin{align}
\label{eq:RIP}
(1 - \delta) \norm{\sig}_2^2 \leq \frac{\nbVert}{\nbVertRed} \norm{\Meas \, \sig}_2^2 \leq (1 + \delta) \norm{\sig}_2^2
\end{align}
for all $\sig \in \spann(\Fou_{\nbClass})$ with probability at least $1-\epsilon$ provided that
\begin{align*}
\nbVertRed \geq \frac{6}{\delta^2} \; \cumCoh_{\nbClass}^2 \; \log\left( \frac{k}{\epsilon} \right).
\end{align*}
\end{theorem}

The above theorem presents a sufficient condition on $\nbVertRed$ ensuring that $\Meas$ satisfies the 
restricted isometry property \refeq{eq:RIP}. This condition is required to ensure 
that the solution of \refeq{eq:dec} is an accurate estimation of $\indClus_j$. The above theorem thus indicates 
that sampling $O(\cumCoh_{\nbClass}^2\log{\nbClass})$ features is sufficient to recover the cluster 
indicator vectors. 

For a simple regular graph $\Graph$ made of $\nbClass$ 
disconnected clusters, we have seen that $\Fou_\nbClass = (\indClus_1, \ldots, \indClus_\nbClass)$ up to 
a normalisation of the vectors. Therefore,  
$\cumCoh_{\nbClass} = \nbVert^{1/2}/\min_i \{ \nbVert_i^{1/2} \}$, where $\nbVert_i$ is 
the size of the $i^\th$ cluster. If the clusters have the same size $\nbVert_i=\nbVert/\nbClass$ then 
$\cumCoh_{\nbClass} = \nbClass^{1/2}$, the lower bound on $\cumCoh_{\nbClass}$. 
In this simple optimal scenario, sampling 
$O(\cumCoh_{\nbClass}^2\log{\nbClass})=O(\nbClass\log{\nbClass})$ features is thus sufficient to recover the cluster 
indicator vectors. 

The attentive reader will have noticed that for graphs where $\cumCoh_{\nbClass}^2 \approx \nbVert$, no 
downsampling is possible. Yet, a simple solution exists in this situation: variable density sampling. 
Indeed, it is proved in \cite{Puy_ARXIV2015} that, whatever the graph $\Graph$, there always exists an 
optimal sampling distribution such that $\nbVertRed = O(\nbClass\log{\nbClass})$ samples are sufficient to 
satisfy Eq.~\refeq{eq:RIP}. This distribution depends on the profile of the local cumulative coherence and can be 
estimated rapidly (see \cite{Puy_ARXIV2015} for more details). In this paper, 
we only consider uniform sampling to simplify the explanations, but keep in mind that in practice results will always be improved if one 
uses variable density sampling. Note also that one cannot expect to sample less 
than $k$ nodes to find $k$ clusters. Up to the extra $\log(k)$, our result is optimal.


%
\section{CSC in practice}

We have detailed the two fundamental theoretical notions supporting our algorithm, presented 
in Alg.~\ref{alg:CSC}. However, some steps in Alg.~\ref{alg:CSC} still need to be clarified. 
In particular, Sec.~\ref{subsec:non_ideal_filtering} provides an extension of Theorem~\ref{th:id_case_filt} 
that takes into account the use of a non-ideal low-pass filter (to handle the practical case where the order 
of the polynomial approximation is finite). This theorem \textit{in fine} explains and justifies step 4 of Alg.~\ref{alg:CSC}. 
Then, in Sec.~\ref{subsec:choices}, important details are discussed such as 
the estimation of $\eig_\nbClass$ (step 1) and the choice of 
the polynomial approximation (step 2). 
We finish this section with complexity considerations.

\subsection{The CSC algorithm}
\label{sec:algo}

\begin{algorithm}[tb]
   \caption{Compressive Spectral Clustering}
   \label{alg:CSC}
\begin{algorithmic}
%
 \STATE \textbf{Input:} The Laplacian matrix $\Lap$, the number of clusters $\nbClass$; and parameters typically set to $\nbVertRed=2\nbClass\log{\nbClass}$, $\nbFeat=4\log{\nbVertRed}$, $\ordPA=50$ and $\reg=10^{-3}$. 
\vspace{2mm}
 \STATE $\bm{1\cdot}$ Estimate $\Lap$'s $k$-th eigenvalue $\eig_\nbClass$ as in Sec.~\ref{subsec:choices}.
 \STATE $\bm{2\cdot}$ Compute the polynomial approximation $\tilde{h}_{\eig_\nbClass}$ of order $\ordPA$ of the ideal low-pass filter $h_{\eig_\nbClass}$.
 \STATE $\bm{3\cdot}$ Generate $\nbFeat$ random Gaussian signals of mean $0$ and variance $1/\nbFeat$: 
 $\RV=\left(\rv_1|\rv_2|\cdots|\rv_{\nbFeat}\right)\in\Rbb^{\nbVert\times\nbFeat}$. 
 \STATE $\bm{4\cdot}$ Filter $\RV$ with 
 $\ma{\tilde{H}}_{\eig_\nbClass}=\tilde{h}_{\eig_\nbClass}(\Lap)$ as in Sec.~\ref{subsec:fast_filt} 
 and define, for each node $i$, its feature vector $\tilde{\feat}_i\in\Rbb^\nbFeat$:
 \begin{equation*}
  \tilde{\feat}_i= \left[{\left(\ma{\tilde{H}}_{\eig_\nbClass}\RV\right)^\adjoint\vec{\delta}_i}\right] \Big/ {\norm{\left(\ma{\tilde{H}}_{\eig_\nbClass}\RV\right)^\adjoint\vec{\delta}_i}}.
 \end{equation*}
 \STATE $\bm{5\cdot}$ Generate a random sampling matrix $\Meas\in\Rbb^{\nbVertRed\times\nbVert}$ as in Eq.~\eqref{eq:subsampling_matrix_def} and keep only 
 $\nbVertRed$  feature vectors:
 %
$(\tilde{\feat}_{\omega_1}| \ldots|  \tilde{\feat}_{\omega_\nbVertRed})^\adjoint=\Meas(\tilde{\feat}_{1}| \ldots|  \tilde{\feat}_{\nbVert})^\adjoint.$
 %
 \STATE $\bm{6\cdot}$  Run $k$-means on the reduced dataset with the Euclidean distance: 
  $\tilde{D}_{ij}^r=\norm{\tilde{\feat}_{\omega_i}-\tilde{\feat}_{\omega_j}}$
to obtain  $\nbClass$ reduced indicator vectors $\indClus_j^r \in \Rbb^{\nbVertRed}$, one for each cluster.
 \STATE $\bm{7\cdot}$  Interpolate each reduced indicator vector $\indClus_j^r$ with the optimisation 
 problem of Eq.~\eqref{eq:dec}, 
 to obtain the $\nbClass$ 
 indicator vectors $\tilde{\indClus_j}^* \in \Rbb^{\nbVert}$ on the full set of nodes.  
\end{algorithmic}
\end{algorithm}
%
%


As for SC (see Sec.~\ref{subsec:SP}), the algorithm starts with the adjacency matrix $\ma{W}$ of a 
graph $\Graph$. After computing its Laplacian $\Lap$, the CSC algorithm is summarized in Alg.~\ref{alg:CSC}. 
The output $\tilde{\indClus}_j^*(i)$ is not binary and in fact quantifies how much node $i$ belongs to 
cluster $j$, useful for fuzzy 
partitioning. To obtain an exact partition of the nodes, we normalize each indicator vector $\tilde{\indClus}_j^*$, and assign node $i$ to the cluster $j$ for which $\tilde{\indClus}_j^*(i)/\norm{\tilde{\indClus}_j^*}$ is maximal. 

\subsection{Non-ideal filtering of random signals}
\label{subsec:non_ideal_filtering}

In this section, we improve Theorem~\ref{th:id_case_filt} by studying how the error of the polynomial 
approximation $\tilde{h}_{\eig_\nbClass}$ of $h_{\eig_\nbClass}$ propagates to the spectral distance estimation, 
and by taking into account the fact that $k$-means is performed on the reduced set of features 
$(\tilde{\feat}_{\omega_1}| \ldots|  \tilde{\feat}_{\omega_\nbVertRed})^\adjoint=\Meas(\tilde{\feat}_{1}| \ldots|  \tilde{\feat}_{\nbVert})^\adjoint$. 
We denote by $\Meas\ma{Y}_\nbClass\in\Rbb^{\nbVertRed\times\nbClass}$ the ideal reduced feature matrix. We have
$(\feat_{\omega_1}|\cdots|\feat_{\omega_\nbVertRed})^\adjoint=\Meas(\feat_{1}|\cdots|\feat_{\nbVert})^\adjoint=\Meas\ma{Y}_\nbClass$.  
The actual distances we want to estimate using random signals are thus, for all $(i, j) \in \{1, \ldots, \nbVertRed\}^2$
%
\begin{equation*}
 D_{ij}^r := \norm{\feat_{\omega_i}-\feat_{\omega_j}}=\norm{\ma{Y}_\nbClass^\adjoint\Meas^\adjoint (\vec{\delta}_i^r-\vec{\delta}_j^r)},
\end{equation*}
where the $\{\vec{\delta}_i^r\}$ are here Diracs in $\nbVertRed$ dimensions. 

Consider the random matrix $\RV=\left(\rv_1|\rv_2|\cdots|\rv_\nbFeat\right)\in\Rbb^{\nbVert\times\nbFeat}$ constructed as in Sec.~\ref{subsec:ideal_filtering}. Its filtered, normalized and reduced version is $\Meas\ma{V}_\nbClass\ma{\tilde{H}}_{\eig_\nbClass}\RV\in\Rbb^{\nbVertRed\times\nbFeat}$. The new feature vector $\tilde{\feat}_{\omega_i}\in\Rbb^\nbFeat$ associated to node $\omega_i$ is thus
%
\begin{equation*}
\tilde{\feat}_{\omega_i}=(\Meas\ma{V}_\nbClass\ma{\tilde{H}}_{\eig_\nbClass}\RV)^\adjoint\vec{\delta}_i^r.
\end{equation*}
The normalisation of Step 4 in Alg.~\ref{alg:CSC} approximates the action of $\ma{V}_\nbClass$ in the above equation. 
More details and justifications are provided in the ``Important remark'' at the end of this section. 
The distance between any two features reads 
\begin{equation*}
 \tilde{D}_{ij}^r := \norm{\tilde{\feat}_{\omega_i}-\tilde{\feat}_{\omega_j}}=\norm{\RV^\adjoint\ma{\tilde{H}}_{\eig_\nbClass}^\adjoint\ma{V}_\nbClass^\adjoint\Meas^\adjoint(\vec{\delta}_i^r-\vec{\delta}_j^r)}.
\end{equation*}
We now study how well $\tilde{D}_{ij}^r$ estimates $D_{ij}^r$.

\textbf{Approximation error.} 
Denote $\ePA(\lambda)$ the approximation error of the ideal low-pass filter: 
%
\begin{equation*}
 \forall\lambda\in[0, 2], \qquad \ePA(\lambda):=\tilde{h}_{\lambda_\nbClass}(\lambda)-h_{\eig_\nbClass}(\lambda).
\end{equation*}
In the form of graph filter operators, one has
 \begin{equation*}
\tilde{h}_{\eig_\nbClass}(\Lap)=\ma{\tilde{H}}_{\eig_\nbClass}= \ma{H}_{\eig_\nbClass}+\ma{\EPA}=h_{\eig_\nbClass}(\Lap) + \ePA(\Lap).
 \end{equation*}
%
We model the error $\ePA$ using two parameters: $\ePA_1$ (resp. $\ePA_2$) the maximal error 
for $\lambda\leq\lambda_\nbClass$ (resp.  $\lambda>\lambda_\nbClass$). We have
\begin{equation*}
 \ePA_1 :=\sup_{\lambda \in \{\eig_1, \ldots, \eig_{k}\}} |\ePA(\lambda)|,\qquad \ePA_2 :=\sup_{\lambda\in \{\eig_{k+1}, \ldots, \eig_{\nbVert}\}} |\ePA(\lambda)|.
\end{equation*}
%

\textbf{The resolution parameter.} 
In some cases, the ideal reduced spectral distance $D_{ij}^r$ may be null. 
In such cases, approximating $D_{ij}^r = 0$ using a non-ideal filter is not possible. 
In fact, non-ideal filtering introduces an irreducible error on the estimation of the feature vectors that is 
not possible to compensate in general. 
We thus introduce a resolution parameter $D_{min}^r$ below which the distances $D_{ij}^r$ do not need to 
be approximated exactly, but should remain below $D_{min}^r$ (up to a tolerated error). 
%
%

%
\vspace{2mm}
\begin{theorem}[General norm conservation theorem]
\label{thm:norm}
Let $D_{min}^r \in \left]0,\sqrt{2}\right]$ be a chosen resolution parameter. 
For any $\delta \in\, ]0,1]$, $\beta>0$, if $\nbFeat$ is larger than
\begin{align*}
\frac{16(2+\beta)}{\delta^2-\delta^3/3}\log{\nbVertRed},
\end{align*}
then, for all $(i,j)\in \{1, \ldots, \nbVertRed\}^2$,
\begin{equation*}
(1-\delta) D_{ij}^r\leq\tilde{D}_{ij}^r\leq (1+\delta)D_{ij}^r,
\; \text{ if } \; D_{ij}^r\geq D_{min}^r,
\end{equation*}
and
\begin{equation*}
\tilde{D}_{ij}^r< (1+\delta)D_{min}^r,
\; \text{ if } \; D_{ij}^r<D_{min}^r,
\end{equation*}
with probability at least $1-2\nbVertRed^{-\beta}$ provided that
\begin{align}
\label{eq:max_norm_e2}
\sqrt{\abs{\ePA_1^2 - \ePA_2^2}} \; + \; \frac{\sqrt{2} \; \ePA_2}{D_{min}^{r}\min_i \{v_\nbClass(i)\}}  \; \leq \;  \frac{\delta}{2+\delta}.
\end{align}
\end{theorem}
%
The proof is provided in the 
supplementary material. 

\textbf{Consequence of Theorem \ref{thm:norm}.} All distances smaller (resp. larger) than the chosen resolution 
parameter $D_{min}^r$ are estimated smaller than $(1+\delta)D_{min}^r$ (resp. correctly estimated up to a relative  
error $\delta$). 
Moreover, for a fixed distance estimation error $\delta$, the lower we decide to 
fix $D_{min}^r$, the lower should also be the errors $\ePA_1$ and/or $\ePA_2$ to ensure that 
Eq.~\eqref{eq:max_norm_e2} still holds, which implies an increase of the order $p$ of the 
polynomial approximation of the ideal filter $h_{\eig_\nbClass}$, and ultimately, that means a higher computation time 
for the filtering operation of the random signals. 

\textbf{Important remark.} The feature matrix $\ma{V}_\nbClass\ma{\tilde{H}}_{\eig_\nbClass}\RV$ can be easily computed 
if one knows the cut-off value $\eig_\nbClass$ and the local cumulative coherences 
$v_\nbClass(i)$. Unfortunately, this is not the case in practice. We propose a solution to estimate 
$\eig_\nbClass$ in Sec.~\ref{subsec:choices}. To estimate $v_\nbClass(i)$, one can use the results in Sec.~4 of~\cite{Puy_ARXIV2015} showing that $v_\nbClass(i)=\norm{\Fou_\nbClass^\adjoint\vec{\delta}_i} \approx \norm{(\ma{H}_{\eig_\nbClass}\RV)^\adjoint\vec{\delta}_i}$. 
Thus, a practical way to estimate $\ma{V}_\nbClass\ma{\tilde{H}}_{\eig_\nbClass}\RV$ is to first compute 
$\ma{\tilde{H}}_{\eig_\nbClass}\RV$ and then normalize its rows to unit length, as done in Step 4 of Alg.~\ref{alg:CSC}. 

\subsection{Polynomial approximation and  
estimation of $\eig_\nbClass$}
\label{subsec:choices}

\noindent\textbf{The polynomial approximation.} Theorem~\ref{thm:norm} uses a separate control on 
$\ePA(\lambda)$ below $\eig_\nbClass$ (with $\ePA_1$) and above $\eig_\nbClass$ (with $\ePA_2$). To have such 
a control in practice, one would need to 
use rational filters (ratio of two polynomials) to approximate $h_{\eig_\nbClass}$. Such filters have been 
introduced in the graph 
context~\cite{shi2015infinite}, but they involve another optimisation step that would burden our main message. 
We prefer to simplify our analysis by using polynomials for which only the maximal error can be controlled. We write
\begin{equation}
\label{eq:def_etot}
 \ePA_{m} := \max(\ePA_1,\ePA_2) = \sup_{\lambda \in \{\eig_1, \ldots, \eig_\nbVert\}}\abs{\ePA(\lambda)}.
\end{equation}
In this easier case, one can show that Theorem~\ref{thm:norm} is still valid if Eq.~\eqref{eq:max_norm_e2} 
is replaced by
%
\begin{equation}
\label{eq:bound_etot}
 \frac{\sqrt{2} \; \ePA_{m}}{D_{min}^r\min_i \{v_\nbClass(i)\}}\leq\frac{\delta}{2+\delta}. 
\end{equation}
In our experiments, we could follow~\cite{shuman_DCOSS2011} and use truncated Chebychev polynomials to 
approximate the ideal filter, as these polynomials are known to require a small degree to ensure a given 
tolerated maximal error $\ePA_{m}$. We prefer to follow~\cite{Napoli_arxiv2014} who 
 suggest to use Jackson-Chebychev polynomials: Chebychev polynomials to which are added damping 
 multipliers to alleviate the unwanted Gibbs oscillations around the cut-off frequency $\lambda_\nbClass$. 

\textbf{The polynomial's order $\ordPA$.} 
For a fixed $\delta$, $D_{min}^r$, and $\min_i\{v_\nbClass(i)\}$, one should use the Jackson-Chebychev polynomial of smallest order $\ordPA^*$ ensuring that $\ePA_m$ satisfies Eq.~\eqref{eq:bound_etot}, in order to optimize the computation time 
while making sure that Theorem~\ref{thm:norm} applies. Studying $\ordPA^*$ theoretically without 
computing the Laplacian's complete spectrum (see Eq.~\eqref{eq:def_etot}) is beyond the scope of this paper. Experimentally, $\ordPA=50$ yields good results (see 
Fig.~\ref{fig:recovery_vs_params}c).

\textbf{Estimation of $\eig_\nbClass$}.
The fast filtering step is based on the polynomial approximation of $h_{\eig_\nbClass}$, 
which is itself 
parametrized by $\eig_\nbClass$. Unless we compute the first $\nbClass$ eigenvectors 
of $\Lap$, thereby partly loosing 
our efficiency edge on other methods, we cannot know  
the value of $\eig_\nbClass$ with infinite precision. To estimate it efficiently, 
we use eigencount techniques~\cite{Napoli_arxiv2014}: based on low-pass filtering with a cut-off frequency at $\lambda$ of random signals, one 
obtains an estimation of the number of enclosed eigenvalues in the interval $[0,\lambda]$. Starting with $\lambda=2$ and 
proceeding by dichotomy on $\lambda$, one stops the algorithm as soon as the number of enclosed eigenvalues equals $k$. 
For each value of $\lambda$, in order to have a proper estimation of the number of enclosed eigenvalues, we choose to filter 
 $2\log{\nbVert}$ random signals 
with Jackson-Chebychev polynomial approximation of the ideal low-pass filters. 

\subsection{Complexity considerations}
\label{subsec:complexity}
%
The complexity of steps 2, 3 and 5 of Alg.~\ref{alg:CSC} are not detailed 
as they are insignificant compared to the others.
First, note that fast filtering a graph signal costs 
$O(\ordPA\,\# \mathcal{E})$.\footnote{Recall that $p$ is the order of the polynomial filter.} 
Therefore, Step 1 costs $O(\ordPA\,\# \mathcal{E}\log{\nbVert})$ per iteration 
of the dichotomy, and
Step 4 costs $O(\ordPA\,\# \mathcal{E}\log{\nbVertRed})$ (as $\nbFeat=O(\log{\nbVertRed})$). 
Step 7 requires to solve Eq.~\refeq{eq:dec} with the polynomial approximation of 
$g_{\lambda_\nbClass}(\lambda) = 1 - h_{\eig_k}(\lambda)$. When 
solved, \eg, by conjugate gradient or gradient descent, this step costs a fast filtering operation per 
iteration of the solver and for each of the $k$ classes. Step 7 thus costs $O(\ordPA\,\# \mathcal{E}k)$. 
Also, the complexity of $k$-means to cluster $Q$ feature vectors of dimension $r$ into $k$ 
classes is $O( \nbClass Q r)$ per iteration. Therefore, Step 6 with $Q=\nbVertRed$ and 
$r = d = O(\log(\nbVertRed))$ costs $O(\nbClass\nbVertRed\log{\nbVertRed})$. 
CSC's complexity is thus 
$O\left(\nbClass\nbVertRed\log{\nbVertRed}+\ordPA\,\# \mathcal{E}\left(\log{\nbVert}+\log{\nbVertRed}+\nbClass \right)\right).$ 
In practice, we are interested in sparse graphs: $\,\# \mathcal{E}=O(\nbVert)$. Using the fact that $\nbVertRed=O(\nbClass\log{\nbClass})$, CSC's complexity simplifies to
\begin{equation*}
 O\left(\nbClass^2\log^2{\nbClass}+\ordPA\nbVert\left(\log{\nbVert}+\nbClass\right)\right).
\end{equation*}
SC's $k$-means step has a complexity of $O(\nbVert\nbClass^2)$ per iteration. In many  
cases\footnote{Roughly, all cases for which $\nbClass^2>\ordPA(\log{\nbVert}+\nbClass)$.}  
this sole task is more expensive than the CSC algorithm. On top of this, SC has the additional 
complexity of computing the first $\nbClass$ eigenvectors of $\Lap$, for which the cost of ARPACK - 
a popular eigenvalue solver - is $O(k^3+Nk^2)$ (see, \eg, Sec.~3.2 of \cite{chen11}). 

This study suggests that CSC is faster than SC for large $N$ and/or $k$. The above algorithms'  
number of iterations 
are not taken into account as they are difficult to predict theoretically. 
Yet, the following experiments confirm the superiority of CSC over SC in terms of computational time.


\section{Experiments}
\label{sec:results}

\begin{figure*}
a)\includegraphics[width=0.23\textwidth]{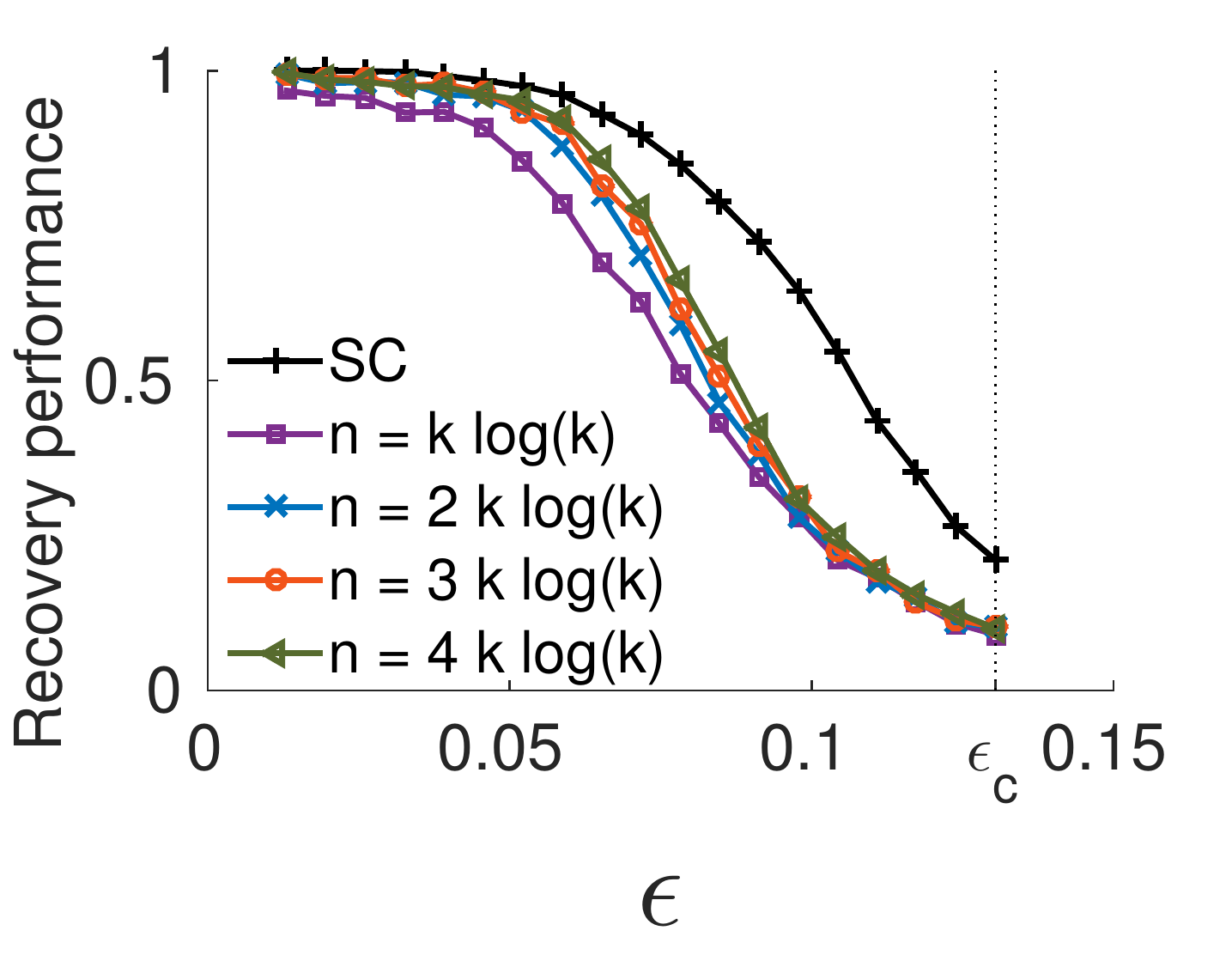}
b)\includegraphics[width=0.23\textwidth]{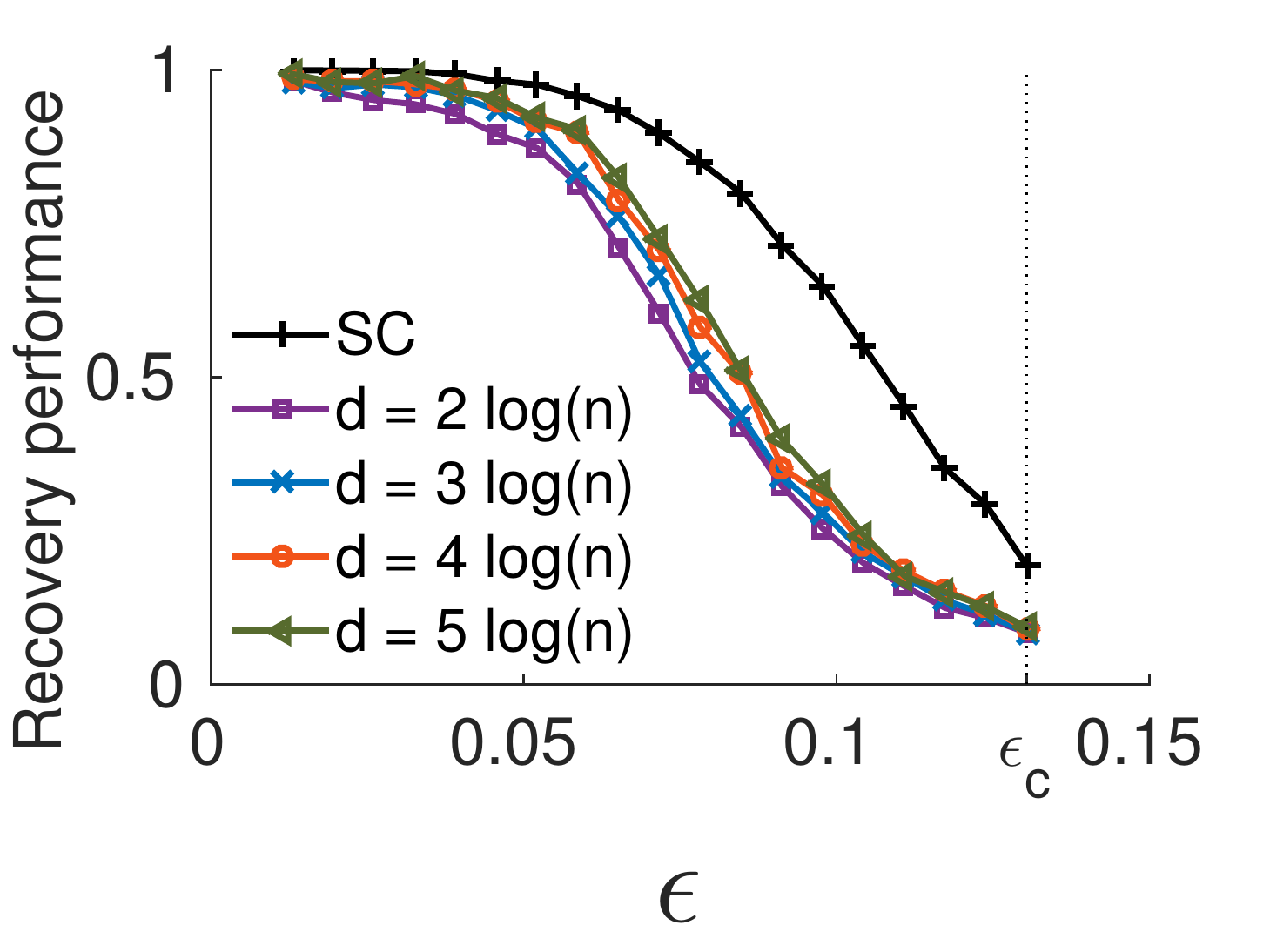}
c)\includegraphics[width=0.23\textwidth]{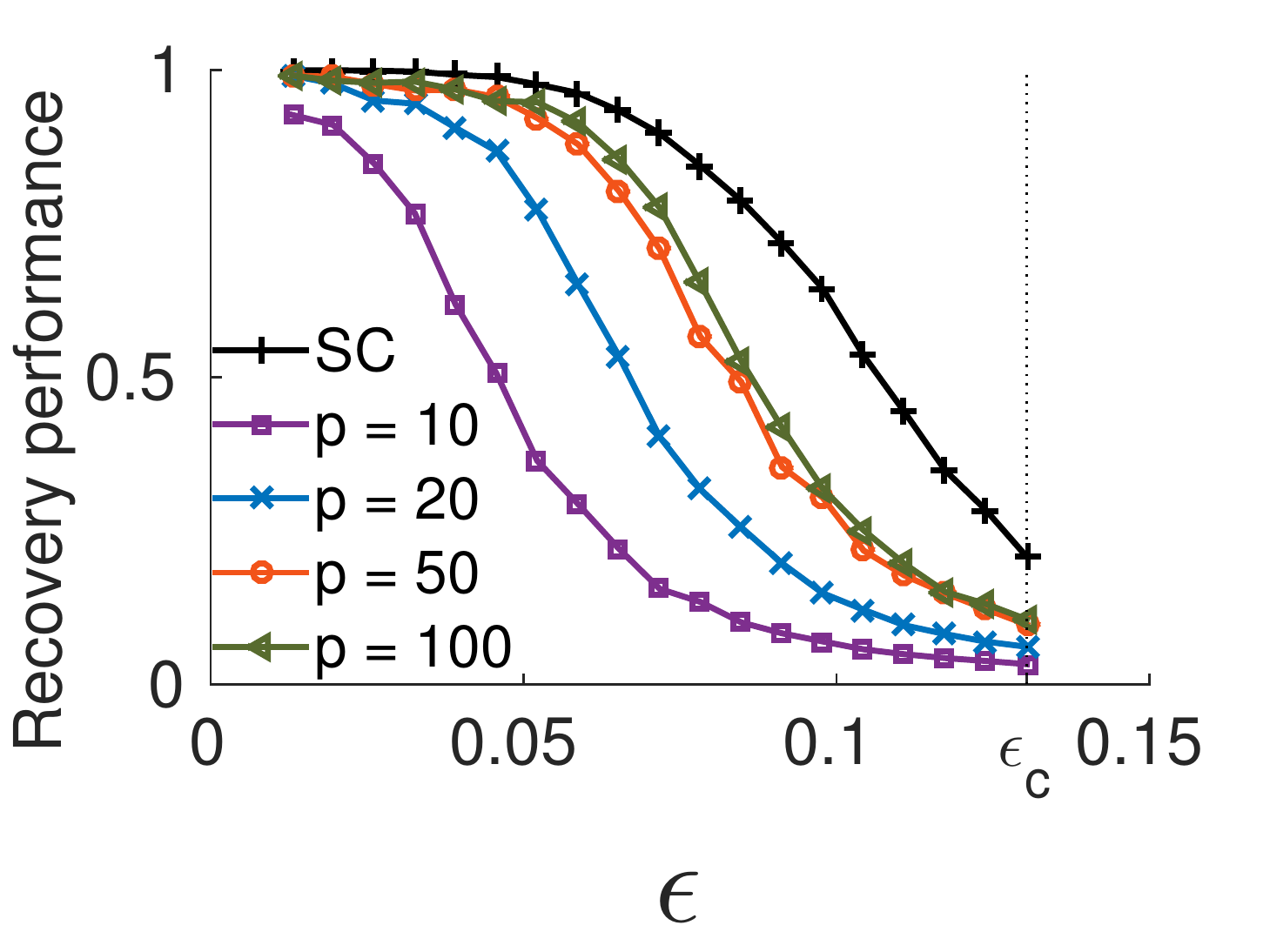}
d)\includegraphics[width=0.23\textwidth]{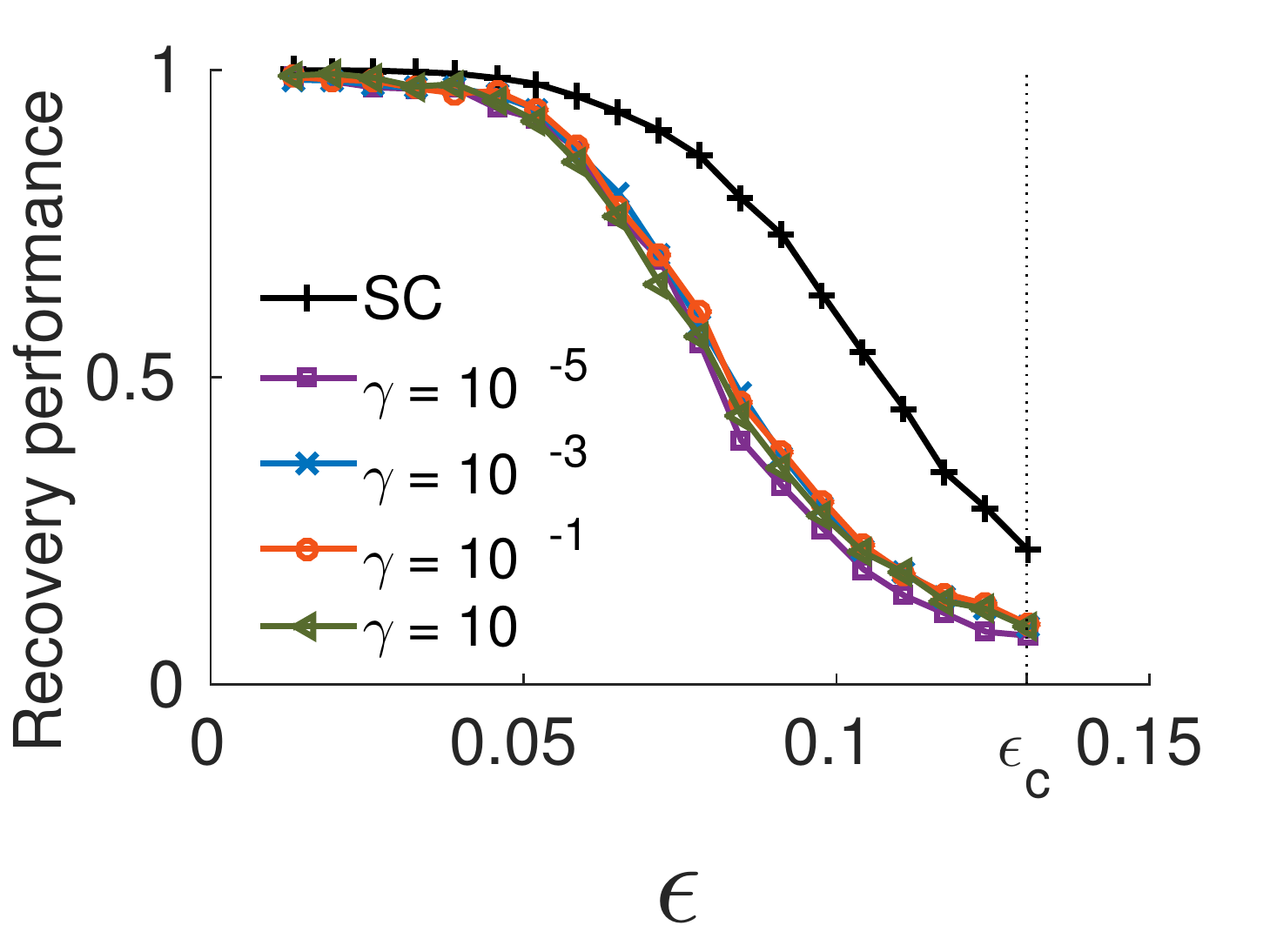}\\
e)\includegraphics[width=0.23\textwidth]{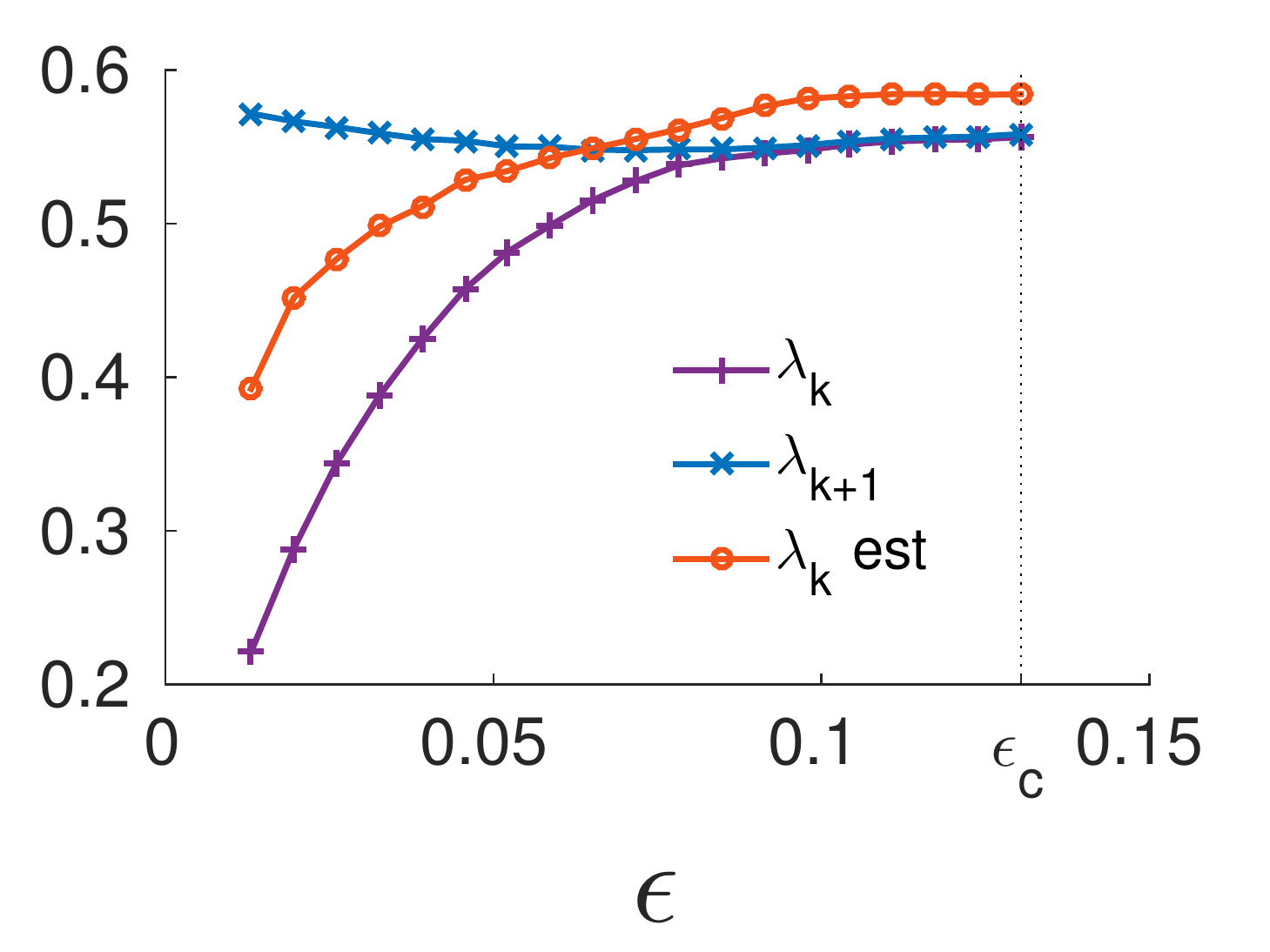}
f)\includegraphics[width=0.26\textwidth]{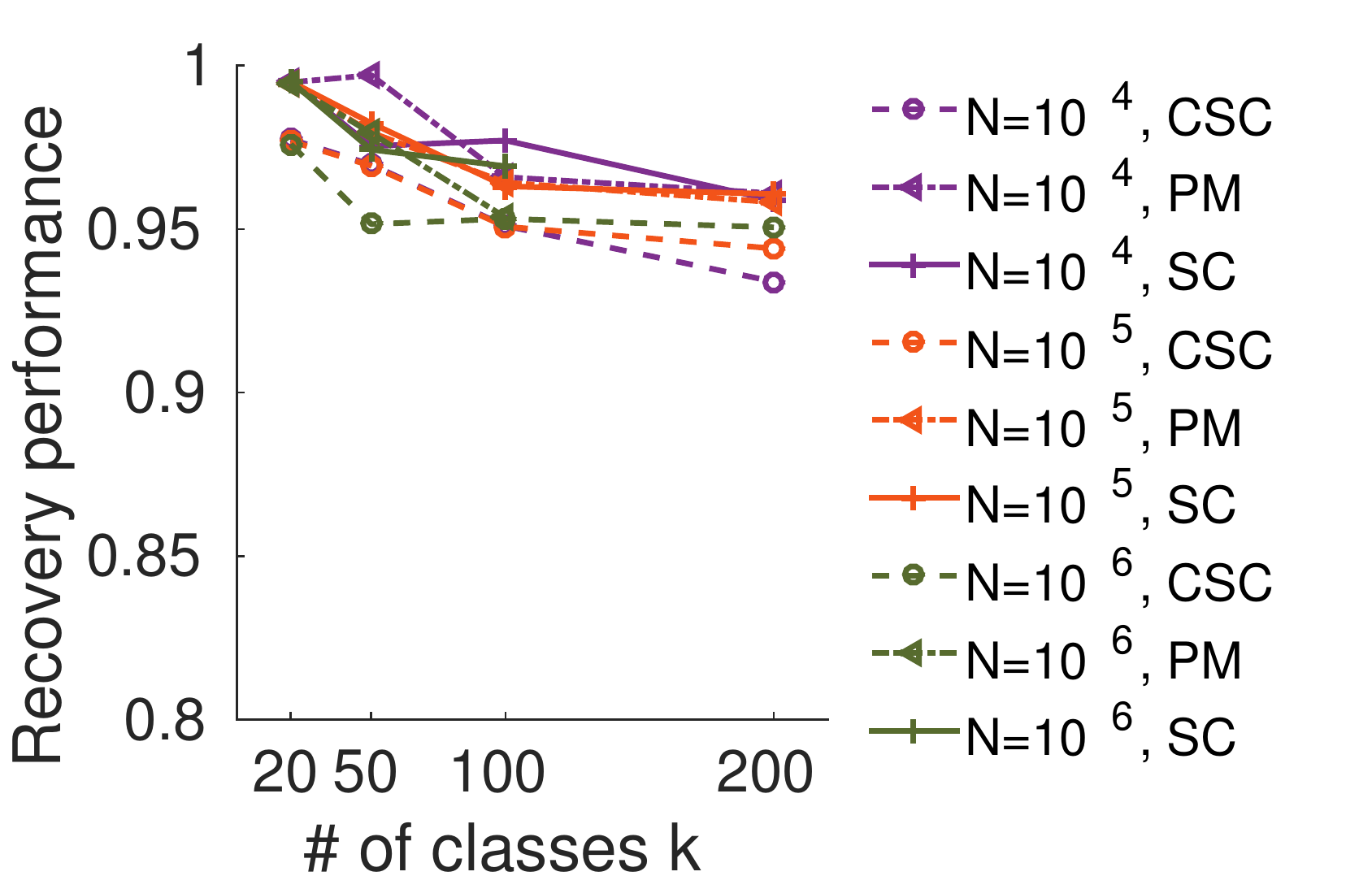}
g)\includegraphics[width=0.18\textwidth]{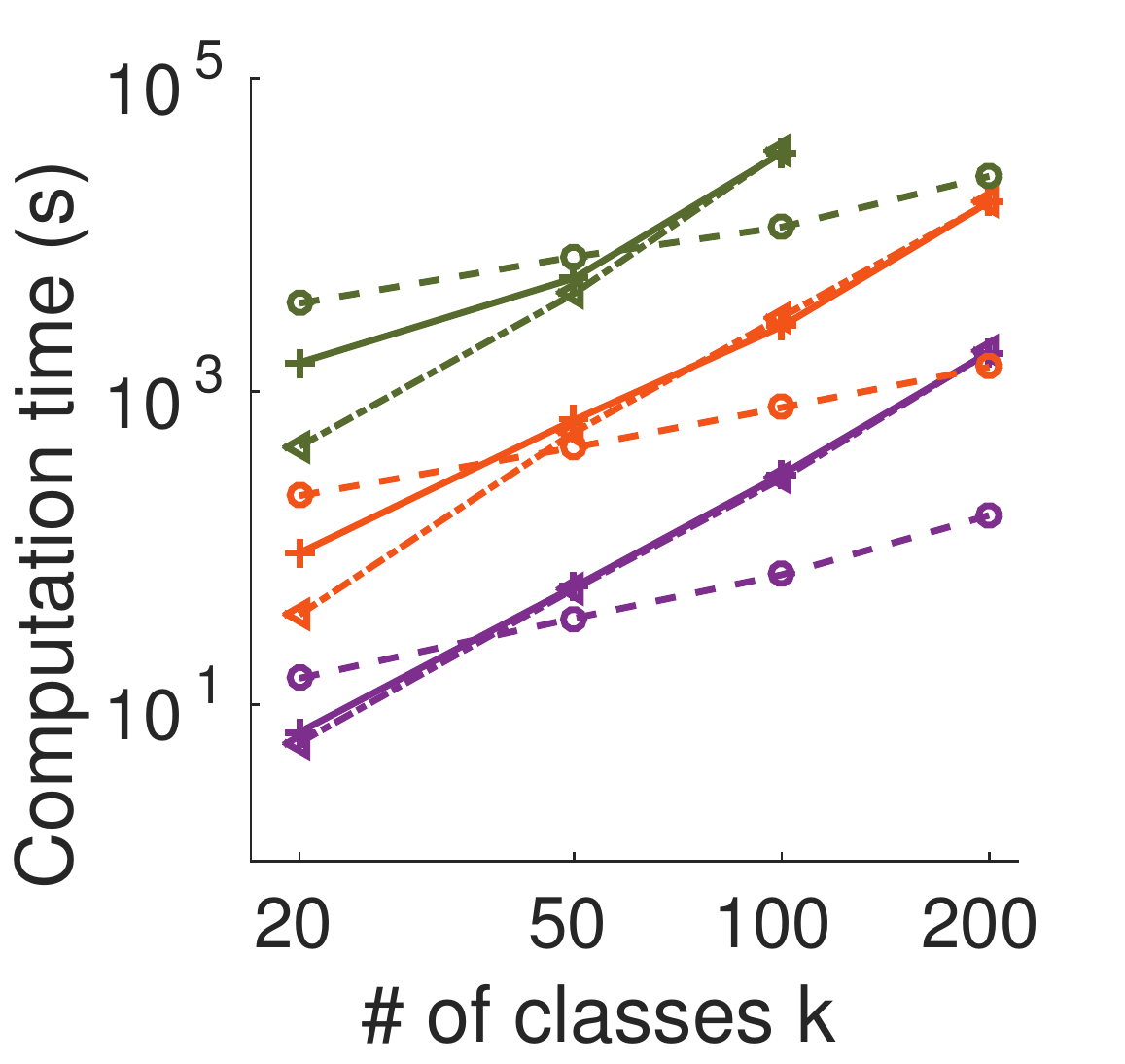}
h)\includegraphics[width=0.23\textwidth]{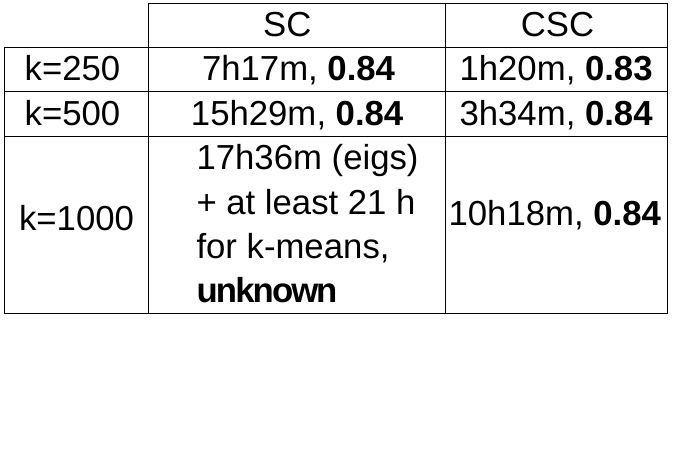}
\caption{(a-d): recovery performance of CSC on a SBM with $\nbVert=10^3, \nbClass=20, s=16$ versus $\epsilon$, 
for different $\nbVertRed$, $\nbFeat$, $\ordPA$, $\reg$. 
Default is $\nbVertRed=2\nbClass\log{\nbClass}$, $\nbFeat=4\log{\nbVertRed}$, $\ordPA=50$ and $\reg=10^{-3}$. 
All results are averaged over 20 graph realisations. 
e)~Estimation of $\lambda_k$ (`$\lambda_k$ est') and the true values of $\lambda_k$ and $\lambda_{k+1}$, 
versus $\epsilon$ on the same SBM. f-g)~Performance and time of computation on a SBM with 
$\epsilon=\epsilon_c/4$ and different values of $\nbVert$ and $\nbClass$; for CSC, PM (Power Method) and SC. For $\nbVert=10^6$ 
and $\nbClass=200$, we stopped SC (and PM) after 20h of computation. Figures f and g are averaged over 3 graph realisations. 
N.B.: Fig.~f is zoomed around high values of the recovery score, and Fig.~g is plotted in log-log. h)~Time of computation (in hours) and modularity (in bold) of the obtained partitions 
for SC and CSC on the Amazon graph. For $\nbClass=1000$, SC's eigendecomposition converges in 17h, 
and we stopped k-means after 21 hours of computation. 
}
\label{fig:recovery_vs_params}
\end{figure*}

We first perform well-controlled experiments on the Stochastic Block Model (SBM), a model of random graphs  
with community structure, that was showed suitable as a benchmark for SC in~\cite{lei_AoStat2015}. 
We also show performance results on a large real-world network. Implementation 
was done in Matlab R2015a, using the built-in function \texttt{kmeans} with 20 replicates, and the function 
\texttt{eigs} for SC. Experiments were done on a laptop with a 2.60 GHz Intel i7 dual-core processor running OS 
Fedora release 22 with 16 GB of RAM. The fast filtering part of CSC uses the \texttt{gsp$\_$cheby$\_$op} function 
of the \texttt{GSP} toolbox~\cite{perraudin2014gspbox}. Equation~\eqref{eq:dec} is solved using 
Matlab's \texttt{gmres} function. All our results are reproducible with the CSCbox downloadable at \url{http://cscbox.gforge.inria.fr/}.

\subsection{The Stochastic Block Model}
What distinguishes the SBM from Erdos-Renyi graphs is that the probability 
of connection between two nodes $i$ and $j$ is not uniform, but depends on the community label of $i$ and $j$. 
More precisely, the probability of connection between nodes $i$ and $j$ 
equals $q_1$ if they are in the same community, and $q_2$ if not. In a first approach, we look at graphs 
with $k$ communities, all of same size $N/k$. 
Furthermore, instead of considering the probabilities, one may fully characterize a SBM by 
providing their ratio $\epsilon=\frac{q_2}{q_1}$, as well as the average degree $s$ of the graph. 
The larger $\epsilon$, the more difficult the community structure's detection. In fact, 
Decelle et al.~\yrcite{Decelle_PRE2011} show that a critical value $\epsilon_c$ exists above which community detection is impossible 
at the large $N$ limit: 
$\epsilon_c=(s-\sqrt{s})/(s+\sqrt{s}(k-1))$. 

\subsection{Performance results}
%
%
In Figs.~\ref{fig:recovery_vs_params}\,a-d), we compare the recovery performance 
of CSC versus SC for different parameters. 
The performance is measured by the Adjusted Rand similarity index~\cite{hubert_Jclassif1985} 
between the SBM's ground truth and the obtained partitions. It varies between $-1$ and $1$. 
The higher it is, the better is the reconstruction. These figures show that the performance of CSC saturates 
at the default values of $\nbVertRed, \nbFeat, \ordPA$ and $\reg$ (see top of Alg.~\ref{alg:CSC}). 
Experiments on the SBM with heterogeneous community sizes are provided in the 
supplementary material and show similar results. 

Fig.~\ref{fig:recovery_vs_params}\,e) shows the estimation results of $\lambda_k$ for different values 
of $\epsilon$ : it is overestimated in the SBM context. As long as the estimated value stays under 
$\lambda_{\nbClass+1}$, this overestimation does not have a strong impact on the method. On the other hand, 
as $\epsilon$ becomes larger than $\sim0.06$, our estimation of $\lambda_\nbClass$ is larger than 
$\lambda_{\nbClass+1}$, 
which means that our feature vectors start to integrate some unwanted information from eigenvectors 
outside of $\spann{(\Fou_\nbClass)}$. Even though the impact of this additional information is 
application-dependent and in some cases insignificant, 
 further efforts to improve the estimation of $\lambda_\nbClass$ would be beneficial to our method. 

In Figs.~\ref{fig:recovery_vs_params}\,f-g) we fix $\epsilon$ to $\epsilon_c/4$, 
$\nbVertRed, \nbFeat, \ordPA$ and $\reg$ to the values given 
in Alg.~\ref{alg:CSC}, and vary $\nbVert$ and $\nbClass$. We compare the recovery performance 
and the time of computation of CSC, SC and Boutsidis' power method~\cite{Boutsidis_ICML2015}. 
The power method (PM), in a nutshell, 1) applies the Laplacian matrix to the power $r$ to $\nbClass$ random signals, 2) computes the left singular 
vectors of the $\nbVert\times\nbClass$ obtained matrix, to extract feature vectors, 3) applies $k$-means in high-dimension (like SC) with these 
feature vectors. In our experiments, we use $r=10$.
The recovery performances are nearly identical in all situations, even though  
CSC is only a few percents under SC and PM (Fig.~f is zoomed around the high values of the recovery 
score). For the time of computation, 
the experiments confirm that all three methods 
are roughly linear in $\nbVert$ and polynomial in $\nbClass$ (Fig.~g is plotted in log-log), with a lower 
exponent for CSC than for SC and PM; such that SC and PM are faster for $\nbClass=20$ but CSC becomes up to an 
order of magnitude faster as $\nbClass$ increases to 200. Note that the SBM is favorable to SC as Matlab's 
function $\texttt{eigs}$ converges very fast in this case, \eg, for $\nbVert=10^5$, it finds the first 
$\nbClass=200$ eigenvectors in less than 2 minutes! 
PM sidesteps successfully the cost of $\texttt{eigs}$, 
but the cost of $k$-means in high-dimension 
is still a strong bottleneck. 

We finally compare CSC and SC on a real-world dataset: the Amazon co-purchasing network~\cite{yang2015defining}. 
It is an undirected connected graph comprising $\nbVert= 334\,863$ nodes and $\#\mathcal{E}=925\,872$ edges. 
The results 
are presented in Fig.\ref{fig:recovery_vs_params}\,h) for three values of $\nbClass$. As there is no clear 
ground truth 
in this case, we use the modularity~\cite{Newman_modularity_2006} to measure the algorithm's clustering 
performance, a well-known cost function that measures how well a given partition separates a network in 
different communities. Note that the 20 replicates of $k$-means would not converge for SC with the default 
maximum number of iterations set to $100$. For a fair comparison with CSC, we used only 2 replicates with a maximum number of 
iterations set to  $1000$ for SC's $k$-means step. We see that for the same clustering performance, CSC is much faster than SC, 
especially as $\nbClass$ increases. The PM algorithm on this dataset does not perform well: even though the 
features are estimated quickly, they apparently do not form clear classes such that its $k$-means step takes even 
longer than SC's. For the three values of $\nbClass$, we stopped the PM algorithm after a time of computation 
exceeding SC's. 


\section{Conclusion}
By graph filtering $O(\log\nbClass)$  random signals, we construct feature vectors 
whose interdistances approach the standard SC feature distances. Then, building upon compressive sensing 
results, we show that one can sample $O(\nbClass\log\nbClass)$ nodes from the set of $\nbVert$ nodes, 
cluster this reduced set of nodes and interpolate the result back to the whole graph. If the low-dimensional 
$k$-means result is correct, \ie, if Eq.~\eqref{eq:hypo} is verified, we guarantee that the interpolation 
is a good approximation of the SC result. 
To improve the clustering result of the reduced set of nodes, 
one could consider the concept of 
community cores~\cite{seifi2013stable}. In fact, as the filtering and the low-dimensional 
clustering steps are fairly cheap to compute, one could repeat these steps for different 
random signals, keep the sets of nodes that are always classified together and use only these 
stable ``cores'' for interpolation.
Our experiments show that 
even without such potential improvements, CSC proves efficient and accurate in synthetic and real-world datasets; 
and could be preferred to SC for large $\nbVert$ and/or $\nbClass$. 


\section{Acknowledgments}
This article was submitted when G. Puy was with INRIA Rennes - Bretagne Atlantique, France. 
This work was partly funded by the European Research Council, PLEASE project (ERC-StG-2011-277906), and by the Swiss National Science Foundation, grant 200021-154350/1 - Towards Signal Processing on Graphs. 

\appendix
\section{Proof of Theorem 3.2}
\label{app:proof_id_case_filt}

\begin{proof}
Note that $\ma{H}_{\lambda_\nbClass}=\Fou_\nbClass\Fou_\nbClass^\adjoint$, and that 
$\ma{Y}_\nbClass=\ma{V}_\nbClass\Fou_\nbClass$. 
We rewrite $\norm{\tilde{\feat}_i-\tilde{\feat}_j}$ in a form that will let us apply the 
Johnson-Lindenstrauss lemma of norm conservation: 
\begin{equation}
\label{eq:norm}
\begin{aligned}
 \norm{\tilde{\feat}_i-\tilde{\feat}_j}&=\norm{\RV^\adjoint\ma{H}_{\lambda_\nbClass}^\adjoint\ma{V}_\nbClass^\adjoint(\vec{\delta}_i-\vec{\delta}_j)}\\
 &=\norm{\RV^\adjoint\Fou_\nbClass\Fou_\nbClass^\adjoint\ma{V}_\nbClass^\adjoint(\vec{\delta}_i-\vec{\delta}_j)}\\
 &=\norm{\RV^\adjoint\Fou_\nbClass(\feat_i-\feat_j)}
\end{aligned}
\end{equation}
where the $\feat_i$ are the standard SC feature vectors. 
Applying Theorem 1.1 of~\cite{Achlioptas_JCSS2003} (an instance of the Johnson-Lindenstrauss lemma) 
to $\norm{\RV^\adjoint\Fou_k(\feat_i-\feat_j)}$, the following 
holds.  If $\nbFeat$ is larger than:
\begin{equation}
 \frac{4+2\beta}{\epsilon^2/2-\epsilon^3/3}\log{\nbVert},
\end{equation}
then with probability at least
  $1-\nbVert^{-\beta}$,
 we have, $\forall(i,j)\in\{1,\ldots,\nbVert\}^2$:
\begin{equation}
\label{eq:bounding}
(1-\epsilon)\norm{\Fou_k(\feat_i-\feat_j)}\leq \tilde{D}_{ij}\leq (1+\epsilon)\norm{\Fou_k(\feat_i-\feat_j)}.\nonumber
\end{equation}
As the columns of $\Fou_k$ are orthonormal, we end the proof:
\begin{equation}
 \forall(i,j)\in[1,\nbVert]^2\quad\norm{\Fou_k(\feat_i-\feat_j)}=\norm{\feat_i-\feat_j}=D_{ij}.\nonumber
\end{equation}
\end{proof}

%
\section{Proof of Theorem 4.1}
\label{app:proof_norm_thm}

\begin{proof} 
Recall that:
$
 \tilde{D}_{ij}^r := \norm{\tilde{\feat}_{\omega_i}-\tilde{\feat}_{\omega_j}}=\norm{\RV^\adjoint\ma{\tilde{H}}_{\eig_\nbClass}^\adjoint\ma{V}_\nbClass^\adjoint\Meas^\adjoint\vec{\delta}_{ij}^r},
$
where $\vec{\delta}_{ij}^r=\vec{\delta}_i^r-\vec{\delta}_j^r$. Given that
$\ma{\tilde{H}}_{\lambda_\nbClass} = \ma{H}_{\lambda_\nbClass} + \EPA$ and using the triangle 
inequality in the definition of $\tilde{D}_{ij}^r$, we obtain
\begin{align}
\label{eq:start_proof}
& \norm{ \RV^\adjoint \ma{H}_{\lambda_\nbClass}^\adjoint \ma{V}_{\nbClass}^\adjoint \Meas^\adjoint\vec{\delta}_{ij}^r } \; -\nonumber \\
& \hspace{3mm} \norm{ \RV^\adjoint \EPA^\adjoint \ma{V}_{\nbClass}^\adjoint \Meas^\adjoint\vec{\delta}_{ij}^r } 
\leq \tilde{D}_{ij}^r 
\leq \norm{ \RV^\adjoint \EPA^\adjoint \ma{V}_{\nbClass}^\adjoint \Meas^\adjoint \vec{\delta}_{ij}^r } \; + \\
& \hspace{45mm} \norm{ \RV^\adjoint\ma{H}_{\lambda_\nbClass}^\adjoint  \ma{V}_{\nbClass}^\adjoint \Meas^\adjoint \vec{\delta}_{ij}^r }, \nonumber
\end{align}
We continue the proof by 
bounding $\norm{\RV^\adjoint\ma{H}_{\lambda_\nbClass}^\adjoint \ma{V}_{\nbClass}^\adjoint \Meas^\adjoint\, \vec{\delta}_{ij}^r}$ 
and $\norm{\RV^\adjoint\EPA^\adjoint \ma{V}_{\nbClass}^\adjoint \Meas^\adjoint\, \vec{\delta}_{ij}^r}$ separately.

Let $\delta \in ]0,1]$. To bound 
$\norm{ \RV^\adjoint \ma{H}_{\lambda_\nbClass}^\adjoint \ma{V}_{\nbClass}^\adjoint \Meas^\adjoint \vec{\delta}_{ij}^r }$, 
we set $\epsilon = \delta/2$ in Theorem~3.2. This proves that if $\nbFeat$ is larger than 
\begin{equation*}
\nbFeat_0=\frac{16(2+\beta)}{\delta^2-\delta^3/3}\log{\nbVertRed},
\end{equation*}
then with probability at least $1-\nbVertRed^{-\beta}$,
\begin{equation*}
\left(1-\frac{\delta}{2}\right) D_{ij}^r
\leq \norm{ \RV^\adjoint \ma{H}_{\lambda_\nbClass}^\adjoint \ma{V}_{\nbClass}^\adjoint \Meas^\adjoint \vec{\delta}_{ij}^r } 
\leq \left(1+\frac{\delta}{2}\right) D_{ij}^r,
\end{equation*}
for all $(i,j) \in \{ 1,\ldots, \nbVertRed \}^2$. To bound $\norm{ \RV^\adjoint \EPA^\adjoint\ma{V}_{\nbClass}^\adjoint \Meas^\adjoint \vec{\delta}_{ij}^r }$, we use Theorem $1.1$ in~\cite{Achlioptas_JCSS2003}. This theorem proves that if $\nbFeat>\nbFeat_0$, then with probability at least $1-\nbVertRed^{-\beta}$,
\begin{equation*}
\norm{\RV^\adjoint \EPA^\adjoint \ma{V}_{\nbClass}^\adjoint \Meas^\adjoint \vec{\delta}_{ij}^r}
\leq  \left(1+\frac{\delta}{2}\right) \norm{ \EPA^\adjoint \ma{V}_{\nbClass}^\adjoint \Meas^\adjoint \vec{\delta}_{ij}^r },
\end{equation*}
for all $(i,j) \in \{ 1, \ldots, \nbVertRed \}^2$. Using the union bound and \refeq{eq:start_proof}, we deduce that, with probability at least $1 - 2\nbVertRed^{-\beta}$,
\begin{align}
\label{eq:prob_bound}
& \hspace{-5mm} \left(1-\frac{\delta}{2}\right) D_{ij}^r  - \left(1+\frac{\delta}{2}\right) \norm{\EPA^\adjoint \ma{V}_{\nbClass}^\adjoint \Meas^\adjoint \vec{\delta}_{ij}^r} \nonumber \\ 
& \hspace{25mm} \leq \quad \tilde{D}_{ij}^r \quad \leq \\ 
& \hspace{10mm} \left(1+\frac{\delta}{2}\right) \norm{\EPA^\adjoint \ma{V}_{\nbClass}^\adjoint \Meas^\adjoint\vec{\delta}_{ij}^r} \; + 
\left(1+\frac{\delta}{2}\right) D_{ij}^r, \nonumber
\end{align}
for all $(i,j) \in \{ 1,\ldots, \nbVertRed \}^2$ provided that $\nbFeat>\nbFeat_0$.

Then, as $\ePA$ is bounded by $\ePA_1$ on the first $\nbClass$ eigenvalues of the spectrum and by $\ePA_2$ on the remaining ones, we have
\begin{align*}
\norm{\EPA^\adjoint \ma{V}_{\nbClass}^\adjoint \Meas^\adjoint\, \vec{\delta}_{ij}^r}^2
& = \norm{\Fou \ePA(\ma{\Lambda}) \Fou^\adjoint \ma{V}_{\nbClass}^\adjoint \Meas^\adjoint\, \vec{\delta}_{ij}^r}^2 \\
& = \norm{\ePA(\ma{\Lambda}) \Fou^\adjoint \ma{V}_{\nbClass}^\adjoint \Meas^\adjoint\, \vec{\delta}_{ij}^r}^2 \\
& = \; \sum_{l=1}^\nbVert \ePA(\eig_l)^2 \, \abs{(\Meas \ma{V}_{\nbClass} \fou_l)^\adjoint \vec{\delta}_{ij}^r}^2 \nonumber \\
& \; \leq \;
\ePA_1^2 \; \sum_{l=1}^\nbClass \abs{(\Meas \ma{V}_{\nbClass} \fou_l)^\adjoint \vec{\delta}_{ij}^r}^2 \nonumber \\
& \hspace{10mm} + \ePA_2^2 \; \sum_{l= \nbClass+1}^\nbVert \abs{(\Meas \ma{V}_{\nbClass} \fou_l)^\adjoint \vec{\delta}_{ij}^r}^2 \nonumber \\
& = \ePA_1^2 \;  \norm{\Fou_\nbClass^\adjoint \ma{V}_{\nbClass}^\adjoint \Meas^\adjoint \vec{\delta}_{ij}^r}^2 \nonumber \\
& + \ePA_2^2 \; \left(\norm{\Fou^\adjoint \ma{V}_{\nbClass}^\adjoint \Meas^\adjoint \vec{\delta}_{ij}^r}^2 - \norm{\Fou_\nbClass^\adjoint \ma{V}_{\nbClass}^\adjoint \Meas^\adjoint \vec{\delta}_{ij}^r}^2 \right) \nonumber \\
& = \; (\ePA_1^2 - \ePA_2^2) \; \norm{\Fou_\nbClass^\adjoint \ma{V}_{\nbClass}^\adjoint \Meas^\adjoint \vec{\delta}_{ij}^r}^2 \nonumber \\
& \hspace{10mm} + \ePA_2^2 \; \norm{\Fou^\adjoint \ma{V}_{\nbClass}^\adjoint \Meas^\adjoint \vec{\delta}_{ij}^r}^2 \nonumber \\
& = (\ePA_1^2 - \ePA_2^2) \; (D_{ij}^r)^2 + \ePA_2^2 \; \norm{\ma{V}_{\nbClass}^\adjoint \Meas^\adjoint \vec{\delta}_{ij}^r}^2 \nonumber \\
& \leq (\ePA_1^2 - \ePA_2^2) \, (D_{ij}^r)^2 + \frac{2 \, \ePA_2^2}{\min_i \{v_\nbClass(i)^2\}}.
\end{align*}
The last step follows from the fact that 
\begin{equation*}
\begin{aligned}
 \norm{\ma{V}_{\nbClass}^\adjoint \Meas^\adjoint\, \vec{\delta}_{ij}^r}^2&=\sum_{l=1}^\nbVert \frac{1}{v_\nbClass(l)^2}\abs{(\Meas^\adjoint\vec{\delta}_{ij}^r)(l)}^2\\
 &= \frac{1}{v_\nbClass(\omega_i)^2}+\frac{1}{v_\nbClass(\omega_j)^2}\,\leq\,\frac{2}{\min_i \{v_\nbClass(i)\}^2} \nonumber
\end{aligned}
\end{equation*}

Define, for all $(i,j) \in \{ 1,\ldots, \nbVertRed \}^2$:
$$\ePA_{ij} := \sqrt{\abs{\ePA_1^2 - \ePA_2^2}} D_{ij}^r + \frac{\sqrt{2} \ePA_2}{\min_i \{v_\nbClass(i)\}}.$$ 
Thus, the above inequality may be rewritten as:
\begin{equation*}
\norm{\EPA^\adjoint \ma{V}_{\nbClass}^\adjoint \Meas^\adjoint\, \vec{\delta}_{ij}^r} \leq \ePA_{ij},
\end{equation*}
for all $(i,j) \in \{ 1,\ldots, \nbVertRed \}^2$, which combined with \refeq{eq:prob_bound} yields
\begin{align}
\label{eq:almost_there}
& \left(1-\frac{\delta}{2}\right) D_{ij}^r  - \left(1+\frac{\delta}{2}\right) \ePA_{ij} \nonumber \\ 
& \hspace{30mm} \leq \quad \tilde{D}_{ij}^r \quad \leq \\ 
& \hspace{35mm} \left(1+\frac{\delta}{2}\right) \ePA_{ij} \; + 
\left(1+\frac{\delta}{2}\right) D_{ij}^r, \nonumber
\end{align}
for all $(i,j) \in \{ 1,\ldots, \nbVertRed \}^2$, with probability at least $1 - 2\nbVertRed^{-\beta}$
provided that $\nbFeat>\nbFeat_0$.

Let us now separate two cases. In the case where $D_{ij}^r \geq D_{min}^r>0$, we have
\begin{align*}
 \ePA_{ij} 
& = \frac{\ePA_{ij}}{D_{ij}^r}D_{ij}^r
= \left( \sqrt{|\ePA_1^2-\ePA_2^2|} + \frac{\sqrt{2}\ePA_2}{D_{ij}^r\,\min_i \{v_\nbClass(i)\}} \right) D_{ij}^r \\
& \leq \left( \sqrt{|\ePA_1^2-\ePA_2^2|} + \frac{\sqrt{2} \ePA_2}{D_{min}^r\,\min_i \{v_\nbClass(i)\}} \right) D_{ij}^r\\
& \leq\frac{\delta}{2+\delta}D_{ij}^r.
\end{align*}
provided that Eq.~(7) of the main paper holds. 
Combining the last inequality with \refeq{eq:almost_there} proves 
the first part of the theorem. 

In the case where $D_{ij}^r < D_{min}^r$, we have
\begin{equation*}
\ePA_{ij} 
< \sqrt{|\ePA_1^2-\ePA_2^2|}D_{min}^r + \frac{\sqrt{2}\,\ePA_2}{\min_i \{v_\nbClass(i)\}}
\leq
\frac{\delta}{2+\delta}D_{min}^r.
\end{equation*}
provided that Eq.~(7) of the main paper holds.  Combining the last inequality with \refeq{eq:almost_there} terminates the proof.
\end{proof}

\section{Experiments on the SBM with heterogeneous community sizes}
We perform experiments on a SBM with  $\nbVert=10^3, \nbClass=20, s=16$ and 
hetereogeneous community sizes. More specifically, the list of community sizes is chosen to be:  
$5$, $10$, $15$, $20$, $25$, $30$, $35$, $40$, $45$, $50$, $50$, $55$, $60$, $65$, 
$70$, $75$, $80$, $85$, $90$ and $95$ nodes. In this scenario, there is no theoretical value of $\epsilon$ over 
which it is proven that recovery is impossible in the large $N$ limit. Instead, we vary $\epsilon$ between $0$ and 
$0.2$ and show the recovery performance results with respect to $\nbVertRed$, $\nbFeat$, $\ordPA$ and $\reg$  in 
Fig.~\ref{fig:recovery_vs_params}. Results are similar to the homogeneous case presented in Fig.~1(a-d) of 
the main paper. 

\begin{figure}
a)\includegraphics[width=0.22\textwidth]{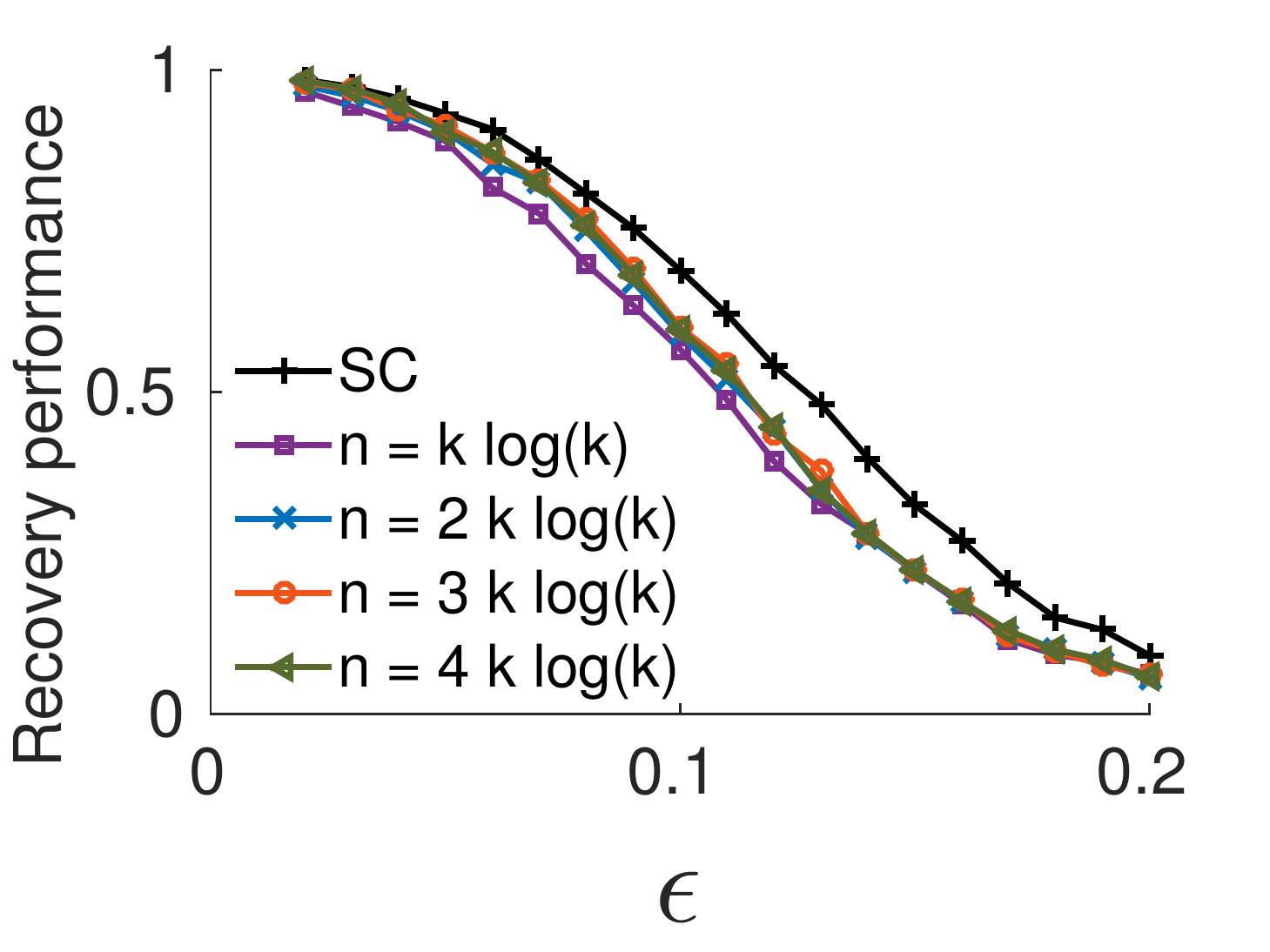}
b)\includegraphics[width=0.22\textwidth]{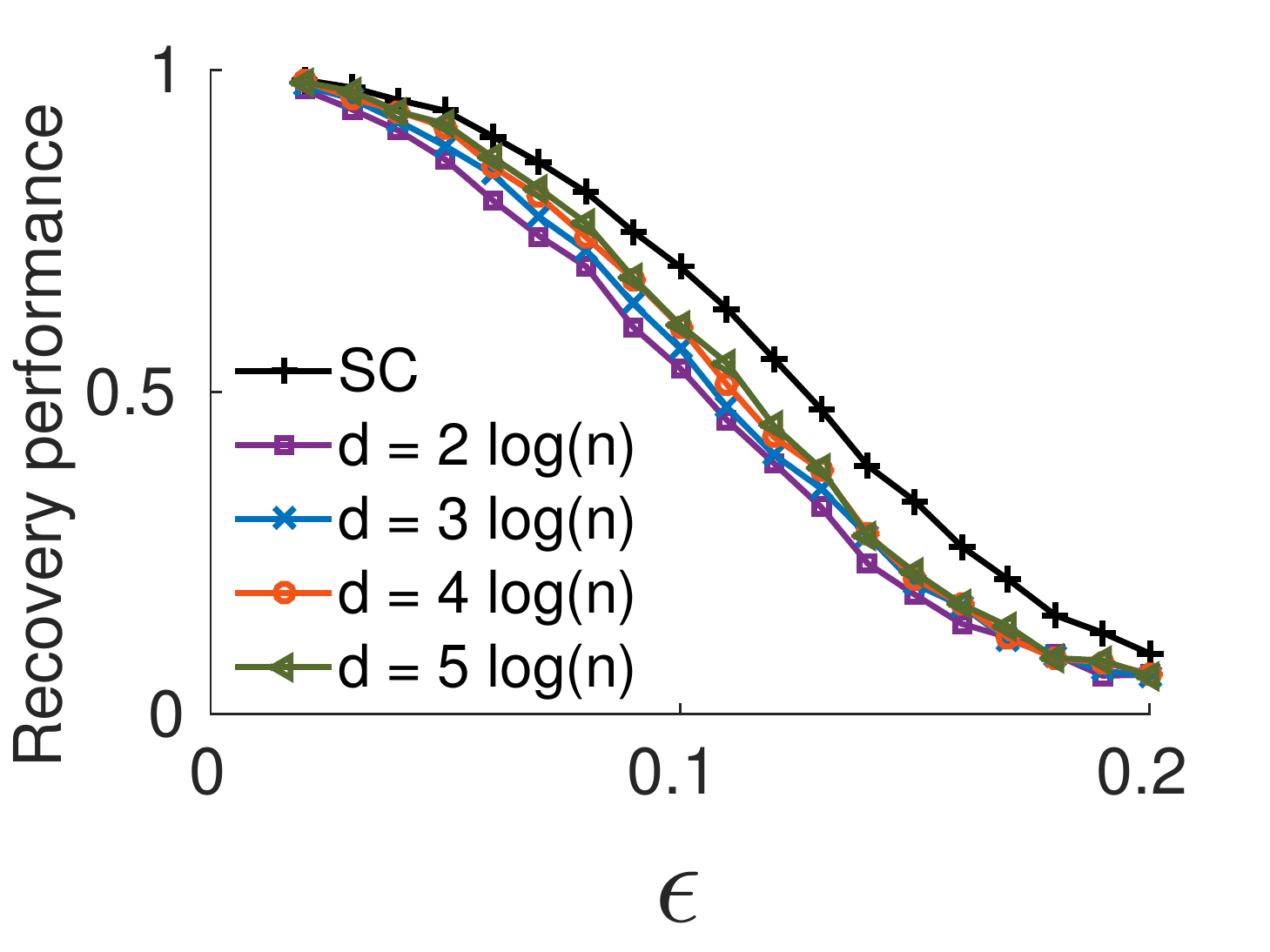}\\
c)\includegraphics[width=0.22\textwidth]{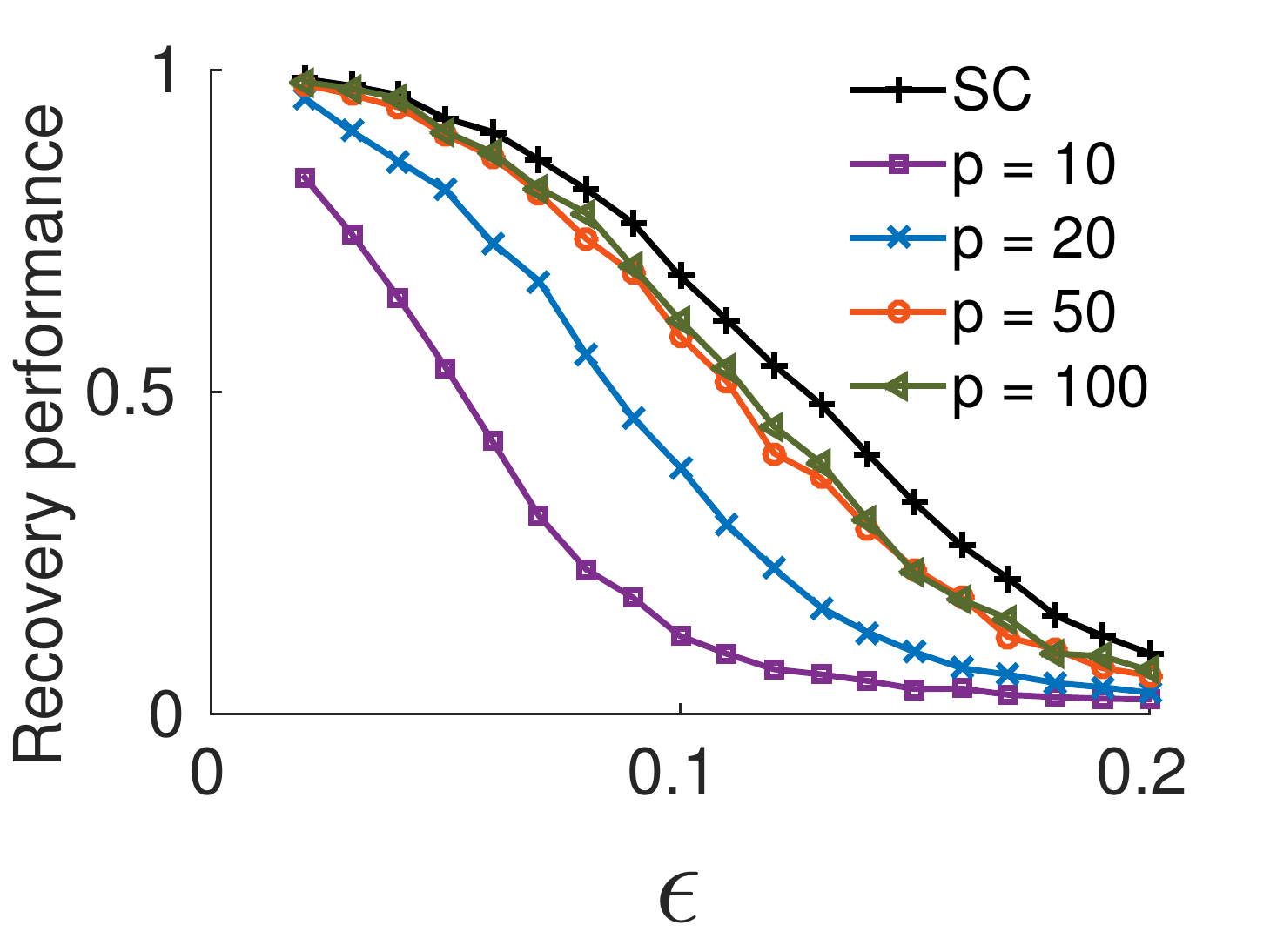}
d)\includegraphics[width=0.22\textwidth]{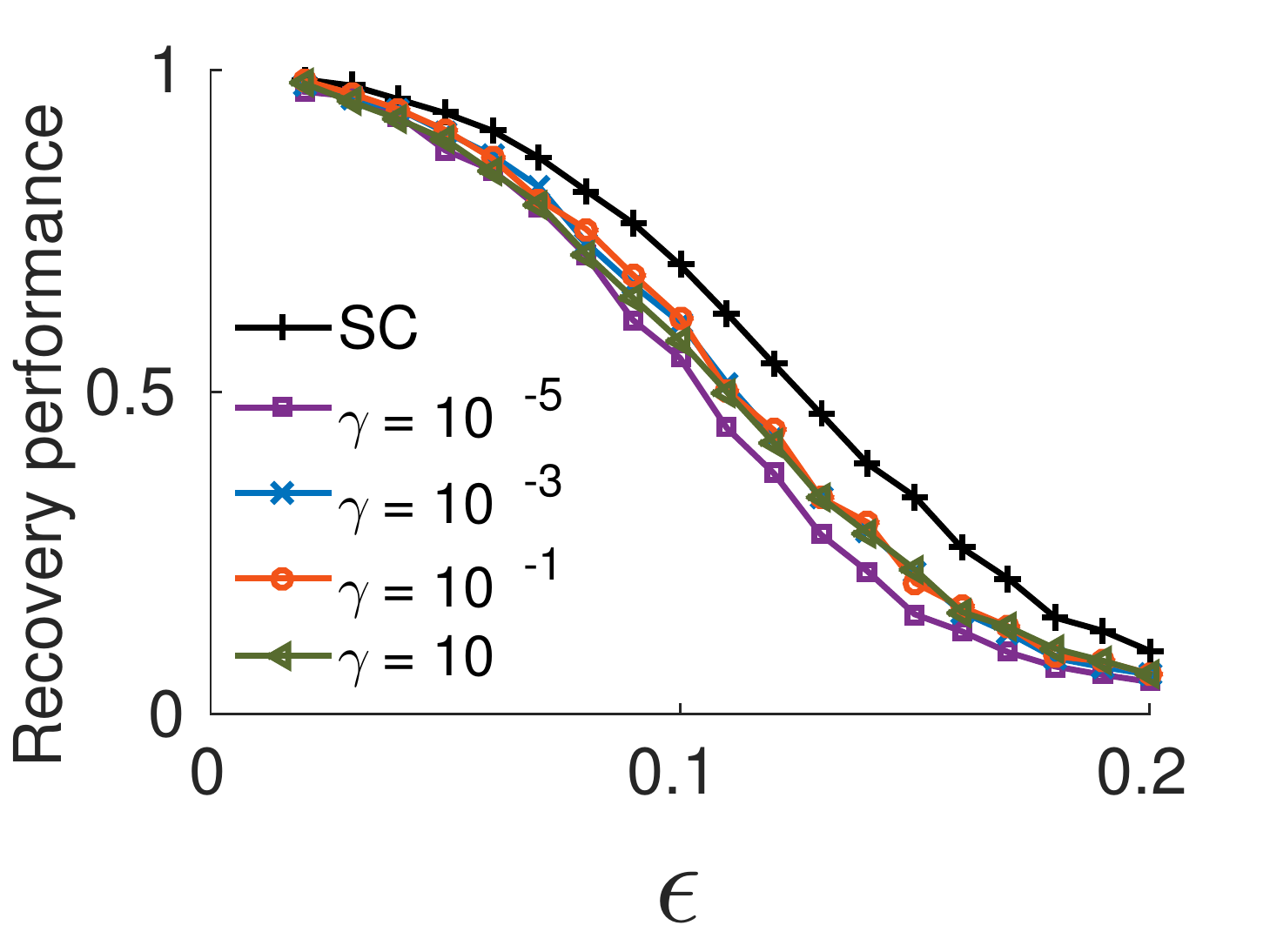}
\caption{(a-d): recovery performance of CSC on a SBM with $\nbVert=10^3, \nbClass=20, s=16$ and 
hetereogeneous community sizes versus $\epsilon$, 
for different $\nbVertRed$, $\nbFeat$, $\ordPA$, $\reg$. 
Default is $\nbVertRed=2\nbClass\log{\nbClass}$, $\nbFeat=4\log{\nbVertRed}$, $\ordPA=50$ and $\reg=10^{-3}$. 
All results are averaged over 20 graph realisations. 
}
\label{fig:recovery_vs_params}
\end{figure}

\bibliography{biblio.bib}
\bibliographystyle{icml2016}

\end{document}

%% file: macro+variables.tex
\newcommand{\Clus}{\ensuremath{\set{C}}}
\newcommand{\Graph}{\ensuremath{\set{G}}}

\newcommand{\Lap}{\ensuremath{\ma{L}}}
\newcommand{\Fou}{\ensuremath{\ma{U}}}
\newcommand{\Eig}{\ensuremath{\ma{\Lambda}}}
\newcommand{\Meas}{\ensuremath{\ma{M}}}
\newcommand{\EPA}{\ensuremath{\ma{E}}}
\newcommand{\RV}{\ensuremath{\ma{R}}}

\newcommand{\rv}{\ensuremath{\vec{r}}}
\newcommand{\fou}{\ensuremath{\vec{u}}}
\newcommand{\indClus}{\ensuremath{\vec{c}}}
\newcommand{\feat}{\ensuremath{\vec{f}}}
\newcommand{\eig}{\ensuremath{\lambda}}

\newcommand{\sig}{\ensuremath{\vec{x}}}

\newcommand{\ePA}{\ensuremath{e}}

\newcommand{\nbVert}{\ensuremath{N}}
\newcommand{\nbVertRed}{\ensuremath{n}}
\newcommand{\nbClass}{\ensuremath{k}}
\newcommand{\nbFeat}{\ensuremath{d}}
\newcommand{\ordPA}{\ensuremath{p}}

\newcommand{\cumCoh}{\ensuremath{\nu}}
\newcommand{\reg}{\ensuremath{\gamma}}

\newcommand{\refeq}[1]{(\ref{#1})}

\newcommand{\Rbb}{\ensuremath{\mathbb{R}}}

\renewcommand{\leq}{\ensuremath{\leqslant}}
\renewcommand{\geq}{\ensuremath{\geqslant}}

\newcommand{\adjoint}{\ensuremath{{\intercal}}}

\newcommand{\norm}[1]{\ensuremath{\left\| #1\right\|}}
\newcommand{\abs}[1]{\ensuremath{\left| #1 \right|}}

\newcommand{\ma}[1]{\ensuremath{\mathsf{#1}}}
\renewcommand{\vec}[1]{\ensuremath{\bm{#1}}}

\newcommand{\set}[1]{\ensuremath{\mathcal{#1}}}

\newcommand{\spann}{\ensuremath{{\rm span}}}

\newcommand{\ie}{\textit{i.e.}}
\newcommand{\eg}{\textit{e.g.}}
\renewcommand{\th}{\ensuremath{\text{th}}}

\newtheorem{theorem}{Theorem}[section]
\newtheorem{definition}[theorem]{Definition}